\documentclass[final,twocolumn]{elsarticle}

\biboptions{sort&compress}

\usepackage{hyperref}
\usepackage{amssymb}
\usepackage{mathtools,amsmath}
\usepackage{cuted}
\usepackage{amsthm}
\usepackage{amsfonts}
\usepackage{amssymb}
\usepackage{graphicx}
\usepackage{dcolumn}
\usepackage{bm}
\usepackage{color}
\usepackage{natbib}

\newcommand{\sslash}{\mathbin{/\mkern-4mu/}}

\journal{Communications in Nonlinear Science and Numerical Simulation}

\DeclareMathOperator{\sech}{sech}
\graphicspath{{./fig-jpg/}{./fig-ps/}{./figures/}{./Figs/}}

\begin{document}

\begin{frontmatter}

\title{Kink-antikink stripe interactions in the two-dimensional sine-Gordon equation}

\author[SDSU]{R. Carretero-Gonz\'{a}lez\corref{cor1}}
\author[Poli]{L.A. Cisneros-Ake}
\author[Hartford]{R. Decker}
\author[UAthens]{G.N. Koutsokostas}
\author[UAthens]{D.J. Frantzeskakis}
\author[UMass]{P.G. Kevrekidis}
\author[Newcastle]{D.J. Ratliff}

\cortext[cor1]{Corresponding author. Email: rcarretero@sdsu.edu}

\address[SDSU]{Nonlinear Dynamical System Group,$^1$\fnref{NLDS}
Computational Science Research Center,$^2$\fnref{CSRC}
and Department of Mathematics and Statistics,
San Diego State University,
San Diego, CA 92182-7720, USA}

\address[Poli]{Departamento de Matem\'aticas, ESFM, Instituto Polit\'{e}cnico Nacional,
Unidad Profesional Adolfo L\'{o}pez Mateos Edificio 9, 07738 Cd.~de
M\'exico, M\'{e}xico}

\address[Hartford]{Mathematics Department, University of Hartford, 
200 Bloomfield Ave, West Hartford, CT 06117, USA}

\address[UAthens]{Department of Physics, National and Kapodistrian University of Athens,
Panepistimiopolis, Zografos, Athens 15784, Greece}

\address[UMass]{Department of Mathematics and Statistics,
University of Massachusetts,
Amherst, MA 01003-4515, USA}

\address[Newcastle]{Department of Mathematics, Physics and Electrical Engineering, Northumbria University, Newcastle-Upon Tyne, NE1 8ST, UK}

\fntext[NLD]{\texttt{URL}: \href{http://nlds.sdsu.edu}{http:$\sslash$nlds.sdsu.edu}}
\fntext[CSRC]{\texttt{URL}: \href{http://www.csrc.sdsu.edu}{http:$\sslash$www.csrc.sdsu.edu}}

\begin{abstract}
The main focus of the present work is to study quasi-one-dimensional kink-antikink 
stripes embedded in the two-dimensional sine-Gordon equation.
Using variational techniques, we reduce the interaction dynamics between 
a kink and an antikink stripe on their respective time and space 
dependent widths and locations. The resulting reduced system of coupled equations
is found to accurately describe the width and undulation dynamics of a
single kink stripe as well as that of interacting ones. 
As an aside, we also discuss two related topics: 
the computational identification of the kink center and its numerical implications 
and alternative perturbative and multiple scales
approaches to the transverse direction induced dynamics for a single kink stripe
in the two-dimensional realm.
\end{abstract}


\end{frontmatter}


\section{Introduction}

Nonlinear Klein-Gordon (KG) models have played a substantial role in the
development of the theory of nonlinear waves and solitons, both at the level of 
completely integrable nonlinear equations, such as the sine-Gordon (sG) model, as well as in the
context of non-integrable models such as the $\phi^4$~\cite{eilbeck,dauxois}.
Indeed, both specialized books~\cite{sgbook,ourphi4} and 
reviews~\cite{kivsharmalomed} have touched upon the relevant topic.
While the non-integrable models, as e.g. the $\phi^4$ model, possess some
intriguing twists, such as the fractal collisions~\cite{Campbell83,anninos,belova}
and the associated irreversible energy transfer phenomena, 
the integrable sG equation has been both a workhorse of 
integrability theory and a ripe testbed of various reduced dynamical theories, such as
collective coordinate ones, among others~\cite{dauxois,kivsharmalomed}.

While these studies have been extensive in the realm of $(1+1)$-dimensions
(one spatial, one temporal) far less has been explored in the context of 
higher dimensional and presumably non-integrable variants.
More specifically, some of the relevant efforts
have focused on the kinematics and dynamics of kinks in two- and three-dimensions 
(2D and 3D)~\cite{christiansen,geicke,bogolub,samuelsen},
considering, among others, their motion driven by curvature and their
ability to produce breathers as a result of collisions with
boundaries~\cite{caputo} and, more recently, their potential towards
stable pinning by local defects~\cite{our_KD_AI}.

Arguably, this lesser focus  may be due to the well-known transverse
stability of the kinks in KG models, contrary to what is the case,
e.g., in their defocusing nonlinear Schr{\"o}dinger
counterparts~\cite{kuzne}, and even in the context of breather generalizations to higher
dimensions; see, e.g., Ref.~\cite{kodama}, as well as the recent revisiting of this
topic in Ref.~\cite{malo}. Nevertheless, we believe it is relevant to
explore the evolution of such ``kink stripes'' in higher dimensions.
This is, on the one hand, due to the relevance of physical realizations of the sG
model in such higher-dimensional settings; see, e.g., Ref.~\cite{boris} for a recent
example. Another motivation is the fact that given its transverse
stability, the relevant kink is a dynamically robust structure whose transverse
dynamics is (also experimentally) relevant to explore. Lastly, this also
paves the way for the exploration of higher-dimensional soliton interactions.
Here, the interplay between the well-known 1D longitudinal kink 
dynamics, which has been long-studied~\cite{Manton} and 
the transverse effects including ones of curvature~\cite{caputo,our_KD_AI}
is naturally of interest to quantify.

It is on that nexus of transverse dynamics and inter-soliton
interaction that the present work will focus. Our corresponding presentation
is structured as follows. Upon discussing the model preliminaries,
in Sec.~\ref{sec:KAK1D}, we present an alternative effective variational
perspective of the kink-antikink (K-AK) interactions 
in 1D which will serve as the basis for building K-AK stripe
reduced models in 2D.
This reduced dimensionality
viewpoint enables us to obtain a quantitative understanding
of the soliton interaction in the 1D setting, and allows
us to quantify concrete features, such as the separatrix between
a pair of K-AK that will periodically re-collide
(being unable to escape each other's attraction) and the
``escape velocity'' scenario, in which they will interact
once and thereafter forever separate. This landscape of K-AK
interactions is first detailed in the $(1+1)$-dimensional setting.
Subsequently, in Sec.~\ref{sec:KAK}, we present the second ingredient
of our considered interplay, namely the transverse dynamics
for a single kink stripe in $(2+1)$-dimensions. Here, again,
the kink center (longitudinal) position and kink width are examined
as a function of time and of the transverse variable $y$.
Also in this section, we bring these two aspects together
and explore the 2D interaction of multiple kinks.
In Sec.~\ref{sec:conclu}, we present our conclusions and forward thoughts
towards how these findings shape and motivate further explorations
along related veins.
In the appendices, we present an alternative perspective of
the two ingredients, namely of the K-AK interaction and the kink stripe transverse
dynamics. The former is via a modified 
diagnostic based on the kink's inflection point, and discusses both
the advantages and disadvantages of adopting such a feature
to identify the coherent structures. The latter is
leveraging a perturbative approach motivated by the work
of Ref.~\cite{pesenson}, unraveling the multiple scales of dynamics along the $y$ direction
of a quasi-1D kink stripe located along the $x$-direction.

\section{Kink-antikink interactions in 1D}
\label{sec:KAK1D}

\subsection{Preliminaries}
\label{sec:KAK1Dprelim}

Before tackling the interactions of quasi-1D kink stripes embedded in 2D,
we first revisit the system of kink-antikink interactions~\cite{eilbeck,dauxois,sgbook,kivsharmalomed} in 1D, which 
will be the foundation towards the generalization for 2D kink stripes~\cite{our_KD_AI}.
The 1D sG equation reads
\begin{equation}
\label{eq:sG1D}
u_{tt}=u_{xx}-V'(u) = u_{xx} - \sin u, 
\end{equation}
where $u(x,t)$ stands for the spatio-temporally
dependent field variable (e.g., the phase difference in the case of superconducting
Josephson junctions or the angular variable in a pendulum array~\cite{sgbook})
and the last term stems from the standard sine-Gordon potential $V(u)=1-\cos(u)$. 
The 1D sG model (\ref{eq:sG1D}) has the associated Lagrangian
\begin{equation}
\label{eq:Lag1D}
L_{\rm 1D}=\int_{-\infty}^{\infty}\left[ \frac{1}{2}u^2_t-\frac{1}{2}u^2_x-(1-\cos u)\right]dx, 
\end{equation} 
and admits kink (K) ($s=1$) and antikink (AK) ($s=-1$) solutions of the form:
\begin{equation}
\label{eq:sGkink1D}
u(x,t)=4\tan^{-1}\left[ \exp\left(sw(x-vt)\right) \right],
\end{equation}
where the Lorentz factor
%
\begin{equation}
\label{eq:w_vs_v}
w=\frac{1}{\sqrt{1-v^2}}, 
\end{equation}
represents the inverse width of the kink with velocity $v$ (with $|v|<1$).
Furthermore, the sG equation also admits exact kink-antikink (K-AK)
symmetric solutions and exact (spatially) localized (temporally)
periodic breather solutions.
As discussed in the above review works
and books and also elsewhere, e.g., more recently in Ref.~\cite{Blas18}, 
the K-AK family corresponds to: 
\begin{equation}
\label{eq:KAK_exact}
u_{\rm KAK}(x,t)=4\tan^{-1} \left[\frac{\sinh(wvt) \sech(wx)}{v}\right],
\end{equation}
where, as above for the single kink, $v$ is the asymptotic velocity for the kinks 
(the kink and antikink travel with equal magnitude but opposite 
velocities) and $w$ is the Lorentz factor given by Eq.~(\ref{eq:w_vs_v}). 
On the other hand, the breather family corresponds to:
\begin{equation}
\label{eq:1D_breather}
u_{\mathrm{b}}(x,t)=4\tan^{-1}\!\left[ {\frac{ \sqrt{1\!-\!\omega^{2}} }{\omega}}
\,{\sech}\left( \sqrt{1\!-\!\omega^{2}} \,x\right) \cos \left( \omega\,t\right)
\right],
\end{equation}
where $0<\omega<1$ is the breather frequency.
Interestingly, these two families are connected to each other by a 
transformation involving the complexification of their arguments~\cite{Blas18}.

\subsection{Variational approach for K-AK interactions}
\label{sec:KAK1DVA}

Although the existence of exact solutions offers a complete picture of kinks and K-AK 
interactions in 1D, as described above, we hereby introduce a variational approach to
describe K-AK dynamics. This will serve as a testbed and a stepping stone for extending 
this methodology to the more interesting case of quasi-1D kink
stripes and their interactions in Sec.~\ref{sec:KAK}.
In particular, the variational approach will allow us to obtain an effective 
attractive potential for the interaction between a K and an AK. This effective
potential serves as a bridge between the K-AK and breather solutions by
approximating the breather solution as a K-AK pair trapped by the confining
portion of the effective potential.

When a K and an AK are placed sufficiently far away from
each other, their shapes can be well approximated by their respective exact
one-kink solutions (\ref{eq:sG1D}). However, as the K and AK profiles
get closer and interact (nonlinearly) through their tails, their shape is deformed and the
solution is then described by the exact K-AK profile (\ref{eq:KAK_exact}).
Therefore, as we would like to obtain equations of motion for the
interaction of K and AK solutions and their stripe extensions in 2D,
one has to rely on approximating their individual shapes. 
For this purpose, we employ a variational approximation (VA) 
to follow these interactions. Our approach relies on the assumption 
that the K and AK shapes remain individually close to the 
profile (\ref{eq:sGkink1D}) but with widths, velocities, and positions 
that need to be adjusted over time. 
Thus, we propose the following (superposition) VA ansatz for a K-AK interaction: 
\begin{equation}
\label{eq:KAKVA1Dsplit}
u(x,t)=4\tan^{-1} e^{g_1}-4\tan^{-1} e^{g_2}, 
\end{equation}
where
\begin{equation}
g_i=w_i\left(t\right)\left(x-\xi_i \left(t\right)\right),
\end{equation}
and the ansatz parameters $w_i$ and $\xi_i$ correspond to the inverse widths and positions 
for each of the kinks. It should be noted that this type of ansatz is
often used when characterizing pairwise kink interactions, e.g.,
through the so-called Manton method~\cite{Manton}.

%
%


Unfortunately, independent widths, and thus velocities, for the K and AK in the VA 
ansatz~(\ref{eq:KAKVA1Dsplit}) are not amenable to obtaining explicit expressions 
for the dynamics of the interacting kinks. Thus, we focus our attention to 
the relevant {\em symmetric} collision case where $w=w_1=w_2$ and 
$\xi=-\xi_1=\xi_2$.
Nevertheless, this essentially amounts to considering the interactions
in the center-of-mass frame.
Namely, the K and AK profiles are symmetric and 
they move in opposite directions at a separation $2\xi$.
In this symmetric case, the K-AK ansatz above can be rewritten as:
\begin{equation}
\label{eqKAK}
u(x,t)=u_{\rm VA}(x,t)=4\tan^{-1} \left[\sinh(w \xi) \sech(wx)\right]. 
\end{equation}
Note that this ansatz is very similar to the exact K-AK solution~(\ref{eq:KAK_exact})
with the subtle difference that the exact solution, in addition to
corresponding to $\xi=v t$, includes a factor of $v$ in the
denominator of the argument of $\tan^{-1}$. 
Including this factor in the VA yields unwieldy complicated equations of motion.
%
Therefore, we opt to keep the ansatz~(\ref{eqKAK}) as it is amenable to simpler
equations of motion and it will also be found to be justifiable a
posteriori (based on the quality of the results as shown below). 

Using the simplified ansatz~(\ref{eqKAK}) in Eq.~(\ref{eq:Lag1D}), yields
the following averaged Lagrangian for the symmetric interaction case:
\begin{equation}
L_{\rm 1D}=\frac{1}{2}N\dot w^2+\frac{1}{2}M\dot \xi^2 
+I\dot w\dot \xi
-{\cal V} ,
\label{Lg1D}
\end{equation}
where
\begin{eqnarray}
N(\xi,w)&=&
\frac{4}{3w^3}\left[\pi^2-\frac{2w\xi(\pi^2-8\xi^2w^2)}{S}\right],
\label{eq:NIMV1}
\\
I(\xi,w)&=&
\frac{32w\xi^2}{S},
\label{eq:NIMV2}
\\
M(\xi,w)&=&16w+\frac{32w^2\xi}{S}, 
\label{eq:NIMV3}
\\
{\cal V}(\xi,w)&=&\frac{4}{w}\sech^2\left(w\xi\right)\left(2w\xi+S\right)\tanh\left(w\xi\right) 
\notag
\\
\label{eq:NIMV4}
&&+8w-\frac{16w^2\xi}{S}, 
\end{eqnarray} 
and
\begin{equation}
S(\xi,w)\equiv \sinh(2w\xi).
\end{equation} 
Note that, for ease of notation, we drop from now on the explicit
mention of the dependence on $\xi$ and $w$ of the functions $N$, $I$, $M$, and ${\cal V}$.
%


The general case where $w$ and $\xi$ are both allowed to depend on time
yields, through the corresponding Euler-Lagrange equation of
the averaged Lagrangian (\ref{Lg1D}), the coupled system:
\begin{eqnarray}
M\ddot \xi + I \ddot w &=& -\frac{1}{2} M_{\xi}\dot \xi^2  -{\cal V}_{\xi}
\label{eq:ODEfull1}
\\
\notag
&& 
+ \left(\frac{1}{2}N_{\xi} - I_w\right)\dot w^2
- M_w \dot \xi\dot w,
\\[2.0ex]
I\ddot \xi + N \ddot w &=& -\frac{1}{2} N_w\dot w^2  -{\cal V}_w
\label{eq:ODEfull2}
\\
\notag
&& 
+ \left(\frac{1}{2}M_w - I_{\xi}\right)\dot \xi^2 
- N_{\xi} \dot \xi\dot w,
\end{eqnarray}
%
%
%
where 
the $\xi$ and $w$ subscripts denote partial derivatives.
%
In an effort to extract as much information as possible 
from the VA approach, let us consider the case where 
the width of the kinks $w$ is time-independent.
In particular, the relation
$w^2(1-\dot \xi^2)=1$ indicates that, for small velocities,
the approximation $w\approx 1$ is valid to second order in $\dot \xi$.
Therefore, let us fix $w=1$ and determine the validity and 
usefulness of the VA approach in this setting.

\begin{figure}[t] 
\begin{center}
\includegraphics[width=0.95\columnwidth]{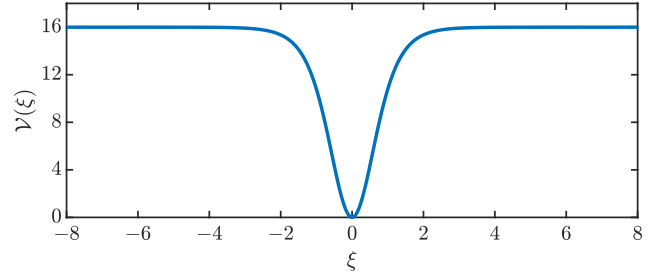}
\caption{Effective attractive potential generated by the kink 
on the anti-kink (and vice-versa).
}
\label{fig:Veff}
\end{center}
\end{figure}

For constant $w$, the Lagrangian (\ref{Lg1D}) reduces to the familiar
form
\begin{equation}
L_{\rm 1D}=\frac{1}{2}M\dot \xi^2-{\cal V},
\label{eq:L1Dwcte}
\end{equation}
where $M$ plays the role of the mass of the effective classical
particle (here the K and AK location) with 
position $\xi$ and ${\cal V}$
corresponding to an effective potential (see below).
Interestingly, for large $\xi$, $M=16$ corresponds to the mass of two
individual kinks (the K and the AK).
The corresponding Euler-Lagrange equations for this case, with $w=1$, yield the following
effective equation of motion of the K-AK position:
\begin{equation}
M\ddot \xi=-\frac{1}{2}\frac{\partial M}{\partial \xi}\dot \xi^2
-\frac{\partial {\cal V}}{\partial \xi}. 
\label{eqmot}
\end{equation} 
Within this setup, the role of ${\cal V}$ is clear, namely it represents the {\em effective}
potential that the K exerts on the AK (and vice-versa). 
%
As depicted in Fig.~\ref{fig:Veff}, the qualitative shape for 
this effective potential lends the first important result of the VA approach.
Namely, the effect of each kink on the other kink is an {\em attractive} 
interaction. Furthermore, this attractive interaction will result in bound 
states where the K and AK will collide with each other periodically if the
energy of the system is below the threshold 
${\cal V}_{\infty}\equiv{\cal V}(\xi\rightarrow\pm\infty)=8(w+1/w)=16$.
For energies above this threshold, the kinks will collide (in forward or backward time)
only once and separate indefinitely.
Precisely at the threshold energy, the system possesses a separatrix between
bounded and unbounded orbits.

\begin{figure}[t] 
\begin{center}
\includegraphics[width=0.95\columnwidth]{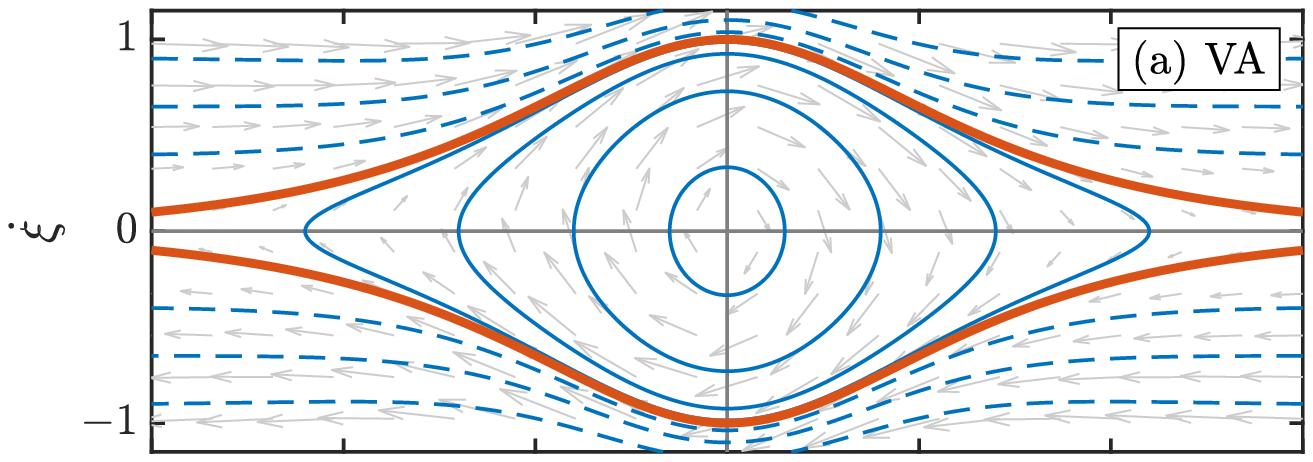}\\
\vspace{0.10cm}
\includegraphics[width=0.95\columnwidth]{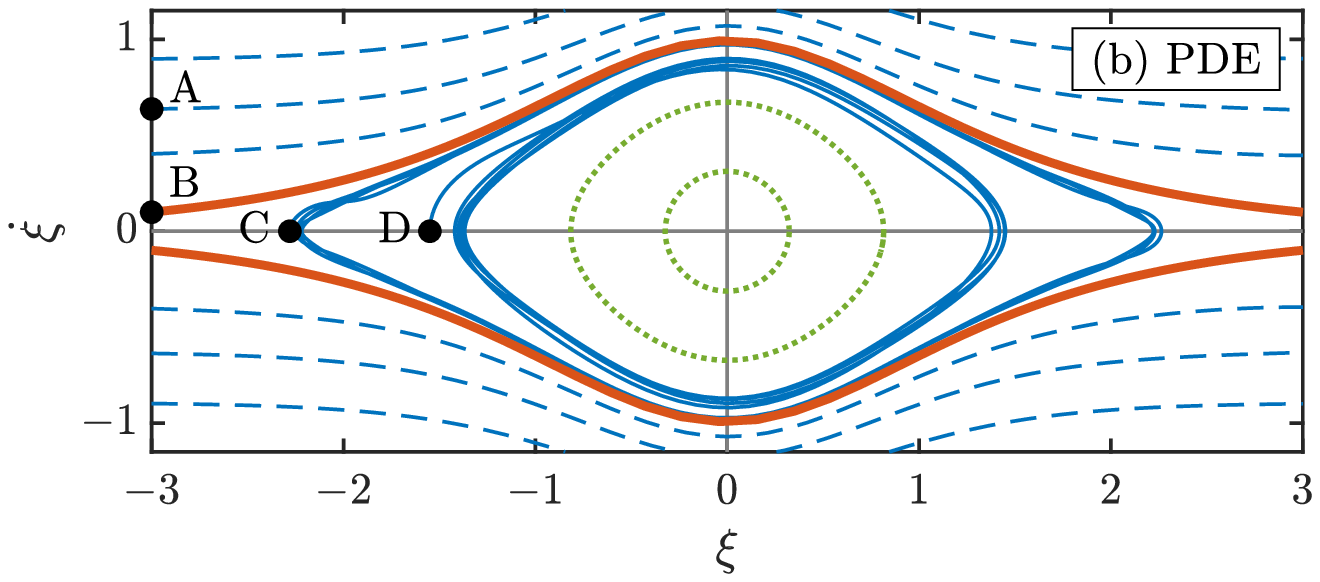}
\caption{(Color online)
Phase plane for the evolution of the kink and anti-kink (half) 
separation distance $\xi$.
The top and bottom panels correspond to the phase planes produced, respectively, by 
(a) the variational approximation reduction of Eq.~(\ref{eqmot}) for $w=1$ and
(b) from direct numerical integration of Eq.~(\ref{eq:sG1D}).
%
In both panels the thick solid (red) lines depict the separatrices separating bound orbits 
(solid and dotted lines) corresponding to periodic K-AK collisional bounces
and unbounded K-AK collision orbits (dashed lines).
For the direct numerical integration case we use the prescription of $\xi$
from Eq.~(\ref{eq:x0FromAnsatz}) and numerically compute its (time) derivative 
using finite differences.
The (green) dotted lines in this case correspond to results obtained from
the exact breathers of Eq.~(\ref{eq:1D_breather}) since the K-AK 
ansatz~(\ref{eqKAK}) fails to reproduce K-AK bounces for such close K-AK 
proximities (cf.~Fig.~\ref{fig:VABR3D}).
The actual evolution of the orbits labeled A through D are depicted in
Fig.~\ref{fig:SGrun_samples}.
}
\label{fig:PhaseSpace}
\end{center}
\end{figure}

\begin{figure}[t] 
\begin{center}
\includegraphics[width=0.95\columnwidth]{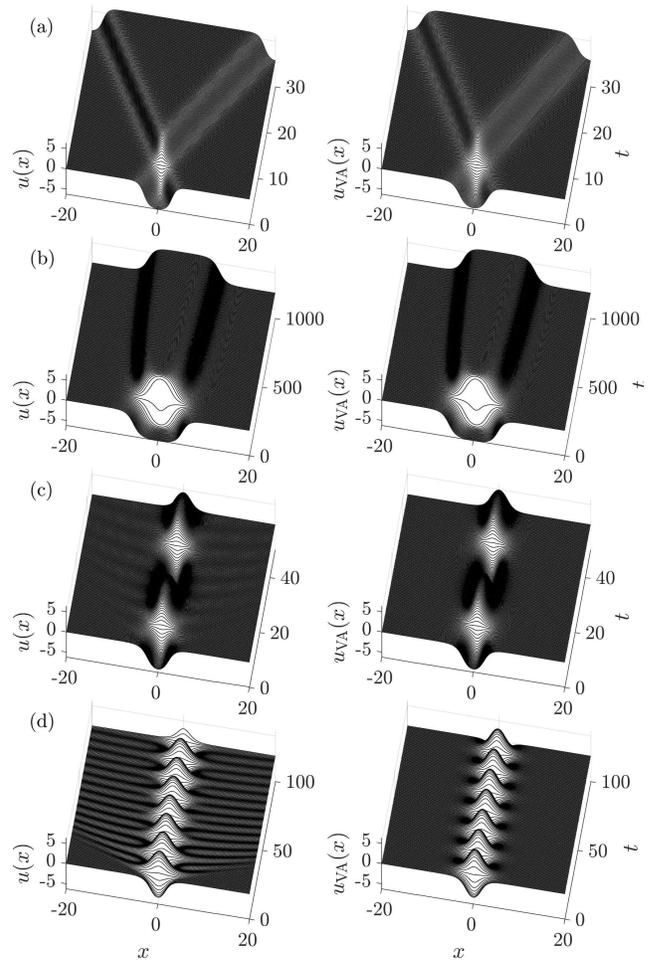}
\caption{
Kink-antikink dynamics.
Left: Evolution according to the full sG model~(\ref{eq:sG1D})
using initial conditions defined by the symmetric K-AK VA ansatz (\ref{eqKAK}).
Right: Corresponding reconstruction of the dynamics using the VA
ansatz (\ref{eqKAK}) with the kink positions determined by the
VA reduced dynamics (\ref{eqmot}).
Rows of panels (a)--(d) correspond to the initial conditions in
phase space denoted by the points A through D in the bottom panel 
of Fig.~\ref{fig:PhaseSpace}.
}
\label{fig:SGrun_samples}
\end{center}
\end{figure}

\begin{figure}[t] 
\begin{center}
\includegraphics[width=0.95\columnwidth]{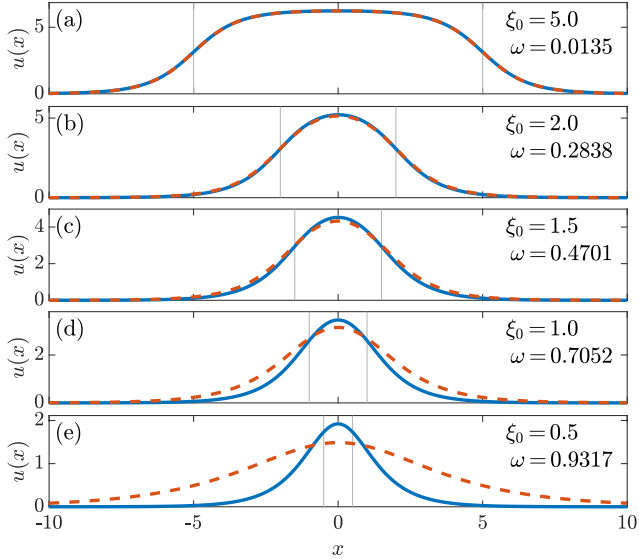}
\caption{(Color online) 
Comparison of the VA ansatz (solid blue line) with  the exact breather
solution (dashed red line) for different K-AK separations $2\xi$.
The corresponding breather frequency $\omega$ was found by fitting
the exact breather profile~(\ref{eq:1D_breather}) to the K-AK VA 
ansatz~(\ref{eq:KAKVA1Dsplit}). The thin vertical lines correspond 
to the positions of the K and AK according to the superposition VA ansatz.
}
\label{fig:VABR1D}
\end{center}
\end{figure}

\begin{figure}[t] 
\begin{center}
\includegraphics[width=0.90\columnwidth]{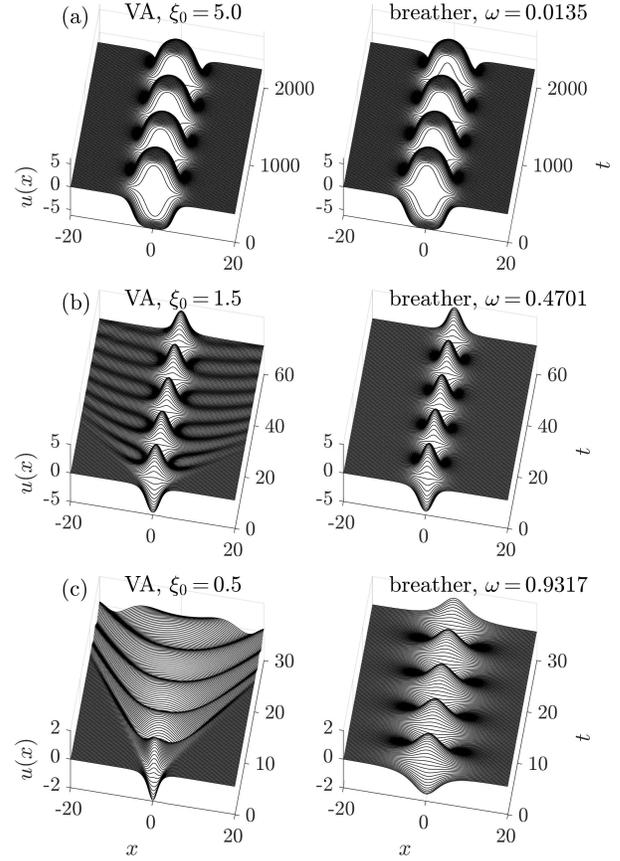}
\caption{Evolution of an initial condition prepared by the
VA ansatz (left column) and the corresponding exact breather
best fit (right column).
The three cases correspond, respectively, to the cases (a), (c), and (e)
of Fig.~\ref{fig:VABR1D}.
(a) For large K-AK separation ($2\xi=10$), the VA profile~(\ref{eq:KAKVA1Dsplit})
approximates very well the exact breather profile~(\ref{eq:1D_breather}).
(b) For a smaller K-AK separation ($2\xi=3$), the VA initial condition captures
qualitatively the essence of the exact breather albeit with large amounts of radiation being shed
during the evolution.
(c) For an even smaller K-AK separation ($2\xi=1$), the VA initial
condition ansatz
cannot even sustain localized oscillations and  is completely destroyed 
during evolution in favor of (spatially) extended oscillations.
}
\label{fig:VABR3D}
\end{center}
\end{figure}

The phase space predicted by the reduced equations of motion (\ref{eqmot}) 
is shown in the top panel of Fig.~\ref{fig:PhaseSpace}.
The bounded and unbounded orbits are depicted, respectively, by the (blue) solid
and dashed lines, while the separatrix is depicted by the thick solid (red) line.
In order to assess the validity of this reduced dynamical system obtained
through the VA methodology, we now reconstruct the phase space from direct
numerical simulations of the original, full sG model (\ref{eq:sG1D}).
A possible method to reconstruct the phase space from the full sG partial differential equation 
(PDE) would be to fit the ansatz to the obtained numerical solution.
However, noting that for $x=0$ the symmetric ansatz (\ref{eqKAK}) yields
$u(0,t)=4\tan^{-1} \left[\sinh(w \xi)\right]$, it is possible to directly
obtain the K-AK separation through:
\begin{equation}
\xi(t) = \frac{1}{w} \sinh^{-1}\left[ \tan\left(\frac{u(0,t)}{4}\right)\right],
\label{eq:x0FromAnsatz}
\end{equation}
by just extracting the time series of the midpoint $u(0,t)$. The bottom panel
of Fig.~\ref{fig:PhaseSpace} depicts the reconstructed phase space from the
full sG numerics for similar initial conditions as the ones used for the
reduced VA phase space depicted in the top panel of the figure.
A sample set of the sG evolutions that were used to create the PDE phase space of
Fig.~\ref{fig:PhaseSpace}, alongside the corresponding VA
  reconstructions of the respective space-time evolutions as obtained
from the ansatz (\ref{eqKAK}), is depicted in Fig.~\ref{fig:SGrun_samples}.
The four rows of panels in Fig.~\ref{fig:SGrun_samples} correspond to
initial conditions depicted by the labels A through D in the bottom (PDE)
panel of Fig.~\ref{fig:PhaseSpace}.
These orbits correspond, respectively, to an unbounded orbit outside of the
separatrix, an orbit on (or, more accurately, very near)
the separatrix itself, and two bounded orbits.
As the figure shows, the reduced VA dynamics (\ref{eqmot}) from the ansatz 
of Eq.~(\ref{eqKAK}) is able to successfully reproduce the original spatio-temporal
K-AK dynamics. Nevertheless, the figure also displays the limitations
of the VA approach in that it is unable to capture the extended wave
radiation that may arise, as shown, e.g., in the comparisons of the
bottom
two rows of the figure.

It is evident that the orbit closer to the
separatrix [Fig.~\ref{fig:SGrun_samples}(c)] does give rise to a  bounded
solution where the K and AK bounce off of each other without any discernible
``radiation''. This is also evident from the extracted phase-space orbit C
in the bottom panel of Fig.~\ref{fig:PhaseSpace}.
However, for the second bounded orbit [Fig.~\ref{fig:SGrun_samples}(d)] the initial
condition provided by the K-AK VA ansatz (\ref{eqKAK}) does no longer produce
an evolution tantamount to solely a periodic bounce of K-AK collisions.
This is also evident from the extracted phase-space orbit D
in the bottom panel of Fig.~\ref{fig:PhaseSpace}, where it is clear that
a sizeable adjustment takes place
through
linear dispersive wavepackets emitted outward toward the boundary
of the domain.
In fact, as the initial separation distance (with zero initial velocity)
between the K and AK is reduced, it is no longer expected that the ensuing 
initial condition prescribed by the superposition K-AK ansatz (\ref{eqKAK})
is valid. This issue is evidenced in the sG evolution corresponding
to $\xi\approx -1.4$ depicted in Fig.~\ref{fig:SGrun_samples}(d) where it 
is clear that the ensuing solution sheds large amounts of radiation
and rapidly deviates from the expected ---for such distances--- bound
state behavior.

Closer inspection of the periodic solutions 
inside the separatrix suggests 
that they resemble the sG breather solutions~(\ref{eq:1D_breather}). 
Figure~\ref{fig:VABR1D} compares the K-AK VA ansatz (\ref{eqKAK}) with the
exact breather profiles given by Eq.~(\ref{eq:1D_breather}) for different
K-AK separation distances. In the comparison, the breather frequency $\omega$ 
is chosen so that the breather profile optimally reproduces (in the least
squares sense) the K-AK VA ansatz.
The figure suggests that a K-AK ansatz profile is able to reproduce
the breather solutions for reasonably large K-AK separations.
In fact, the two inner orbits in the bottom (PDE) panel of Fig.~\ref{fig:PhaseSpace},
depicted by the (green) dotted line, were not produced from K-AK VA initial 
conditions, but they were generated by a breather solution (\ref{eq:1D_breather}) 
for appropriate (small) values of the breather frequency $\omega$.
Figure~\ref{fig:VABR3D} compares the dynamical evolution of an initial
condition seeded with the K-AK VA ansatz (left column of panels) and 
an initial condition seeded with the corresponding breather solution (right
column of panels). As  is evident from the figure, for large separation
distances [see panel (a) for $\xi=5$] the K-AK VA ansatz and breather 
solution are for all practical purposes
equivalent. However, as $\xi$ is progressively reduced the 
K-AK VA ansatz initial condition produces a solution with large amounts
of radiation [see panel (b) for $\xi=1.5$]. Eventually, for small enough
$\xi$, the K-AK VA initial condition is not capable of producing a (spatially)
localized solution and, instead, produces spatially extended oscillations in time
that no longer faithfully mimic the K-AK interaction [see panel (c) for $\xi=0.5$],
with the latter waveform being destroyed, contrary to what we expect
on the basis of the bounded oscillatory orbits of the phase portrait
of Fig.~\ref{fig:PhaseSpace}.

%
From the above results, and in particular the comparison between
the phase portraits in Fig.~\ref{fig:PhaseSpace}, we may
conclude that that the effective VA reduction~(\ref{eqmot}) is indeed able 
to describe the K-AK interactions on {\em both sides} of the separatrix
---not only qualitatively, but also quantitatively--- provided that the separation distance 
between the kinks is larger than about $2\xi\simeq 3$ (i.e., roughly
comparable to twice the width of the kinks). That is to say, as may be expected,
when the separation between the kinks becomes comparable to their width,
the VA can no longer efficiently describe the K-AK interaction.
Nevertheless, the ansatz (\ref{eqKAK}) provides a ``unified'' approach for a
wide range of orbits (and corresponding distances on both sides of
the relevant separatrix) toward 
describing symmetric K-AK interactions. The flexibility provided by this
ansatz will allow us to extend the above 1D K-AK results to the full 2D setting,  
and consider kink stripes and their interactions, as we will see in the next Section.

As a complementary approach, we present in Appendix~\ref{appendix1} 
an alternative perspective on the K-AK interaction dynamics.
This alternative approach is numerically inspired from a modified
kink position diagnostic based on its inflection point, and the
ensuing Newton-type particle interaction equations of motion.

\section{Kink-Antikink Stripes in 2D}
\label{sec:KAK}

\subsection{Single Kink Stripe}

We will leverage the K-AK results from the previous Section for the case
of interacting 2D kink stripes. Let us first obtain effective equations
of motion for a single K (or AK) stripe.
The sine-Gordon equation in the two-dimensional case is:
\begin{equation}
\label{eq:2DsG}
u_{tt}=u_{xx}+\sigma u_{yy}-\sin u, 
\end{equation}
where $u$ stands for the deformation and $\sigma=\pm 1$ reflects the spatial
ellipticity or hyperbolicity of the operator; see, e.g., Ref.~\cite{our_KD_AI}
for a recent consideration of the relevant model.
The numerical results presented
below will chiefly pertain to the `standard' elliptic Laplacian with $\sigma=1$.
However, in our VA methodology below we keep $\sigma$ undetermined as this
will allow the applicability of our findings to be extended to the hyperbolic ($\sigma=-1$) case 
---which includes the potential of a snaking instability of the kink stripe in contrast
with the elliptic case where the stripe is always (neutrally) stable; see below.
In Ref.~\cite{our_KD_AI} reduced, effective, equations of motion
for the undulations of single sG (and general Klein-Gordon) stripe through
an adiabatic invariant (AI) methodology were obtained. However, the
AI approach used therein only allows for a single dependent variable, such as the stripe
center position, to be captured as a function of the transverse coordinate $y$ and time $t$.
Our aim here is to give more freedom to the corresponding manifold of solutions
(i.e., our choice of ansatz solution space) in order to also capture variations
in the width $w$ and the mutual interactions when two such kink stripes interact
in 2D space.
Therefore, we now venture in this direction by describing  the VA methodology 
when applied to an effective 1D kink stripe embedded in 2D space.

The extension of the 1D Lagrangian (\ref{eq:Lag1D}) to 2D for the sG model~(\ref{eq:2DsG}) yields:
\begin{equation}
\label{eq:Lag2D}
L_{\rm 2D}=\int_{-\infty}^{\infty} L_y\,dy, 
\end{equation}
where the averaged (over $x$) Lagrangian density corresponds to:
\begin{equation}
\label{eq:LagLy}
L_y=\int_{-\infty}^{\infty}\left[ \frac{1}{2}u^2_t-\frac{1}{2}u^2_x
-\frac{\sigma}{2}u^2_y+\cos u -1\right]dx. 
~~~~
\end{equation}
We now develop a VA for a kink {\em stripe} of the form:
\begin{equation}
\label{eq4}
u=4\tan^{-1} f\left(x,y,t\right), 
\end{equation}
with
\begin{equation}
\label{eq5}
f=e^{sw\left(y,t\right)\left(x-\xi \left(y,t\right)\right)}=e^{g\left(x,y,t\right)},
\end{equation}
where $s=1$ and $s=-1$ correspond, respectively, to kink and anti-kink stripe.
It is important to mention that this ansatz represents a kink stripe in the 
$y$-direction with (inverse) width $w(y,t)$ and location $\xi(y,t)$ that 
explicitly depend on time and, crucially, on the $y$ coordinate.
Using the trigonometric identities
\begin{equation}
\notag
\cos \theta = \frac{1-\tan^2\frac{\theta}{2}}{1+\tan^2\frac{\theta}{2}}, 
\quad {\rm and} \quad
\tan 2\theta=\frac{2\tan \theta}{1-\tan^2\theta},
\end{equation}
one finds that $1-\cos u=2\sech^2 g$ and $\frac{\partial u}{\partial \alpha}
=2\frac{\partial g}{\partial \alpha}\sech g$ for $\alpha=x,y,t$. 
Then, averaging the Lagrangian (\ref{eq:LagLy}) on the family of trial functions~(\ref{eq4}) 
yields:
\begin{equation}
\label{eq7}
L_y=4sw\left(\xi^2_t-1-\sigma \xi^2_y\right)
+\frac{s\pi^2}{3w^3}\left(w_t^2-\sigma w^2_y\right)-\frac{4s}{w}. ~~~~
\end{equation}
The corresponding equations of motion for the stripe parameters then become:
\begin{eqnarray}
w_{tt}&=&\sigma w_{yy}+\frac{6w^3}{\pi^2}\left(\xi^2_t-1-\sigma \xi^2_y\right)
\notag
\\
\label{eq8}
&&+\frac{3}{2w}\left(w^2_t-\sigma w^2_y\right) +\frac{6w}{\pi^2}, 
\\
\label{eq9}
\xi_{tt}&=&\sigma \xi_{yy}-\frac{1}{w}w_t\xi_t+\frac{\sigma}{w}w_y\xi_y, 
\end{eqnarray}
where we note that the kink's polarity $s$ has disappeared, indicating that 
the dynamics for a kink or antikink stripe are identical.
{%
This Lagrangian approach to transverse dynamics and stability of fronts and solitary waves
has parallels in Whitham modulation theory~\cite{wlnlw,Infeld-Rowlands,kivshar-pelin},
and proves to be more accurate than the classical multiple scales methodology. This methodology
and its connection to the Lagrangian approach can be found in Appendix~\ref{appendix_multiscales}.
}

Before comparing the full stripe dynamics to the reduced VA dynamics of
Eqs.~(\ref{eq8}) and (\ref{eq9}), first we consider the corresponding linear 
stability analysis.
Let us linearize about a co-moving stationary kink (or anti-kink) stripe in a
reference frame at velocity $v_0$ [with its corresponding (inverse) width 
given by Eq.~(\ref{eq:w_vs_v})].
I.e., this is a stripe that travels with speed $v_0$ in the original frame.
Then, perturbing the co-moving stationary state by
$w= w_0+\tilde w$ and $\xi = v_0t+\tilde\xi$, where $|\tilde w|\ll 1$ and
$|\tilde\xi| \ll 1$ are small perturbations, yields
the following dynamical equations for the perturbations:
\begin{eqnarray}
\tilde{w}_{tt}&=&\sigma \tilde{w}_{yy}+\frac{12}{\pi^2}\tilde{w}_0^3v_0\tilde{\xi}_t
-\frac{12}{\pi^2}\tilde{w}, 
\label{eq9a}
\\
\label{eq9b}
\tilde{\xi}_{tt}&=&\sigma \tilde{\xi}_{yy}-\frac{v_0}{w_0}\tilde{w}_t, 
\end{eqnarray}     
where, by construction of the stationary state, $w_0^2\left(1-v_0^2\right)=1$.
Then, considering plane waves of the form $\tilde{w}(t,y)=Ae^{i(ky-\eta t)}$ 
and $\tilde\xi(t,y)=Be^{i(ky-\eta t)}$, of frequency $\eta$ and wave
number $k$, we arrive at the following dispersion relation for
perturbations of wavenumber $k$:
%
\begin{equation}
\label{eq9d}
\eta_{\pm}^2=\sigma k^2+\frac{6}{\pi^2}w_0^2\pm \sqrt{\frac{12}{\pi^2}\sigma k^2\left(w_0^2-1\right)+\frac{36}{\pi^4}w_0^4}. 
~~~~
\end{equation}
Since $0\le v_0 \le 1$ and $w_0^2(1-v_0^2)=1$ one finds that $w_0\ge1$; thus, the last 
expression shows, as expected, that in the elliptic case ($\sigma=1$) one has
full (neutral) stability for the kink stripe for any wavenumber $k$.
On the other hand, in the hyperbolic case ($\sigma=-1$) the kink stripe is generically
unstable and, therefore, we defer its discussion to Appendix~\ref{appendix_hyper}.
From this point onward we focus our results on the elliptic case ($\sigma=1$)
but maintain, for generality purposes, the parameter $\sigma$ in our
model reductions.
%


\begin{figure}[t] 
\begin{center}
\includegraphics[width=0.90\columnwidth]{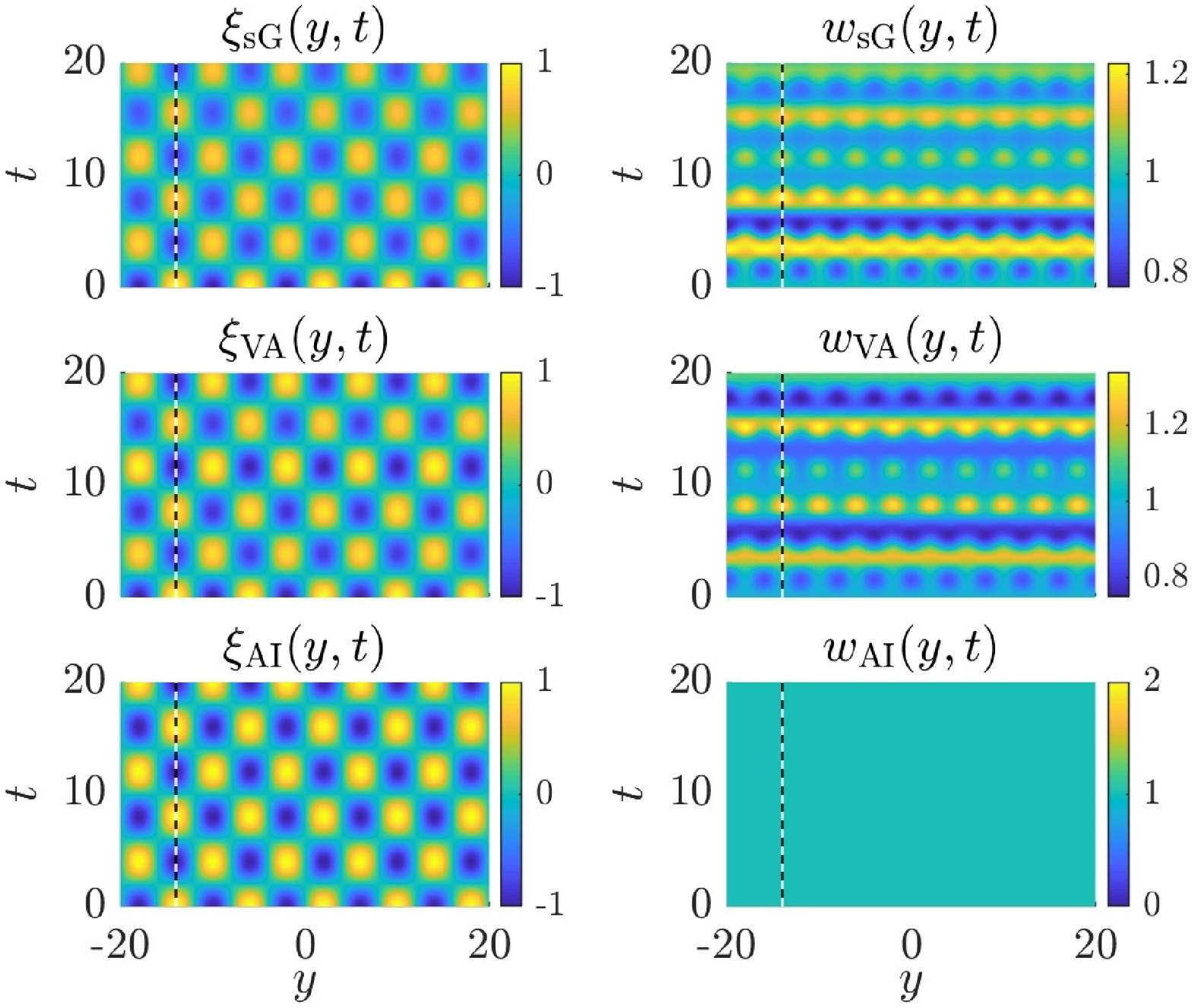} 
\\[1.0ex]
\includegraphics[width=0.80\columnwidth]{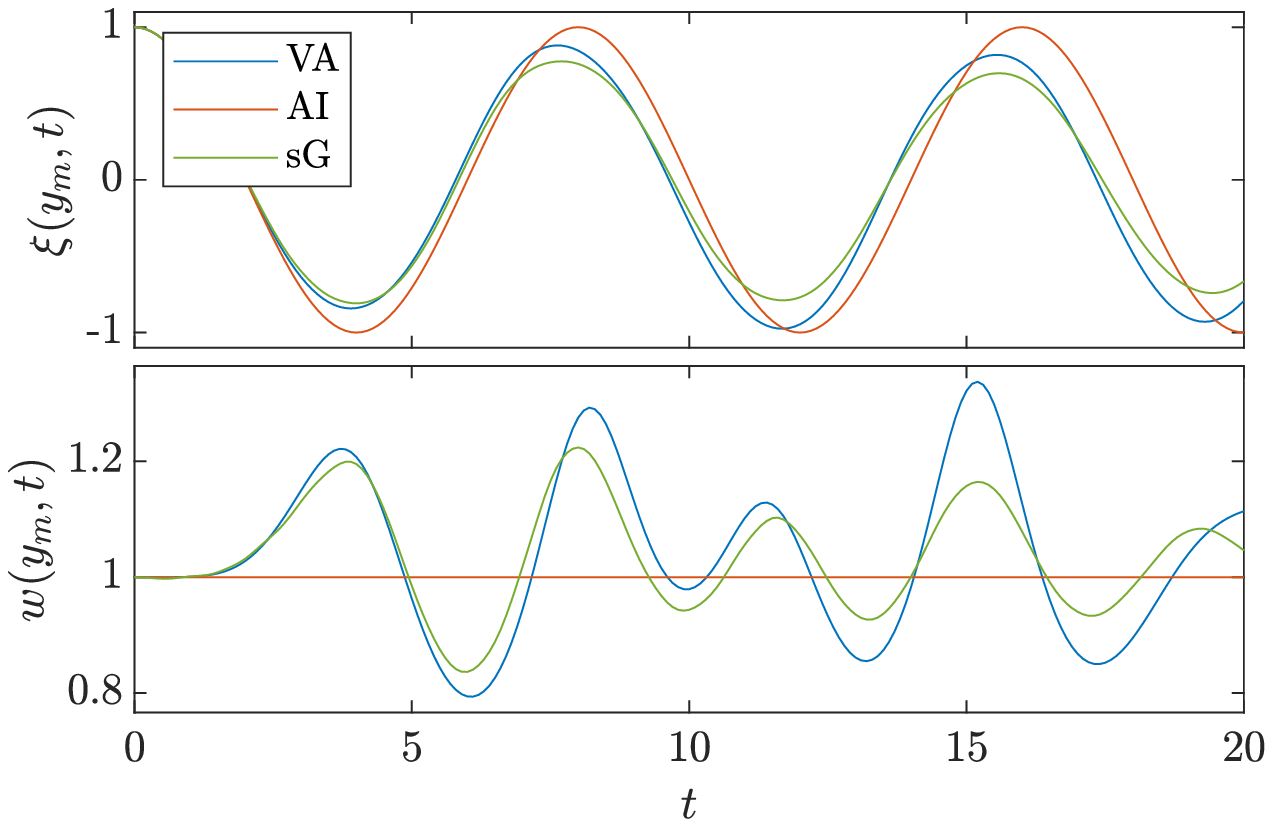}
~
\caption{(Color online)
Spatiotemporal dynamics for a single sG kink stripe. Depicted are the 
position $\xi(y,t)$ and (inverse) width $w(y,t)$ for a stripe initialized with 
an oscillation of the form $\xi(y,0)=\varepsilon \sin(n\pi y/(2\ell))$ and
$\dot\xi(y,0)=w(y,0)=\dot w(y,0)=0$ on the domain $(x,y)\in[-\ell,\ell]\times[-\ell,\ell]$,
with $n=10$, $\ell=20$, and $\varepsilon=1$.
The top panels depict the spatiotemporal $(y,t)$ dynamics for 
$\xi(y,t)$ (left subpanels) and $w(y,t)$ (right subpanels) corresponding to 
the full sG dynamics of Eq.~(\ref{eq:sG1D}) (first row of subpanels),
the stripe VA reduction of Eqs.~(\ref{eq8}) and (\ref{eq9}) (second row of subpanels),
and the AI reduction of Ref.~\cite{our_KD_AI} (third row of subpanels).
The bottom panels depict the corresponding cuts for constant $y=y_m$ at the
location of the first maximum of $\xi(y,0)$ (see vertical dashed lines in the
top subpanels).
%
}
\label{fig:VAAISG2}
\end{center}
\end{figure}

\begin{figure}[t] 
\begin{center}
\includegraphics[width=0.90\columnwidth]{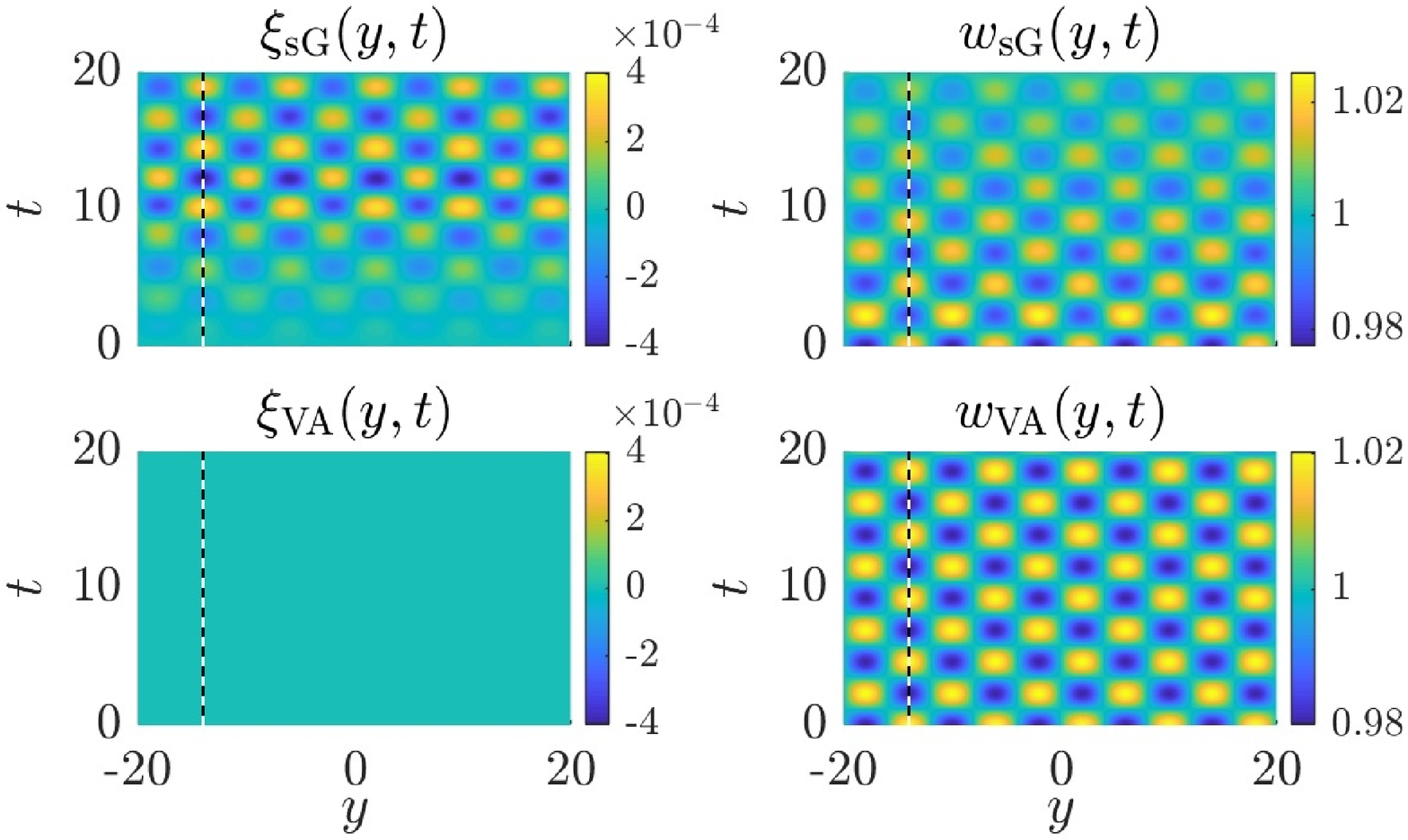}
\\[1.0ex]
\includegraphics[width=0.80\columnwidth]{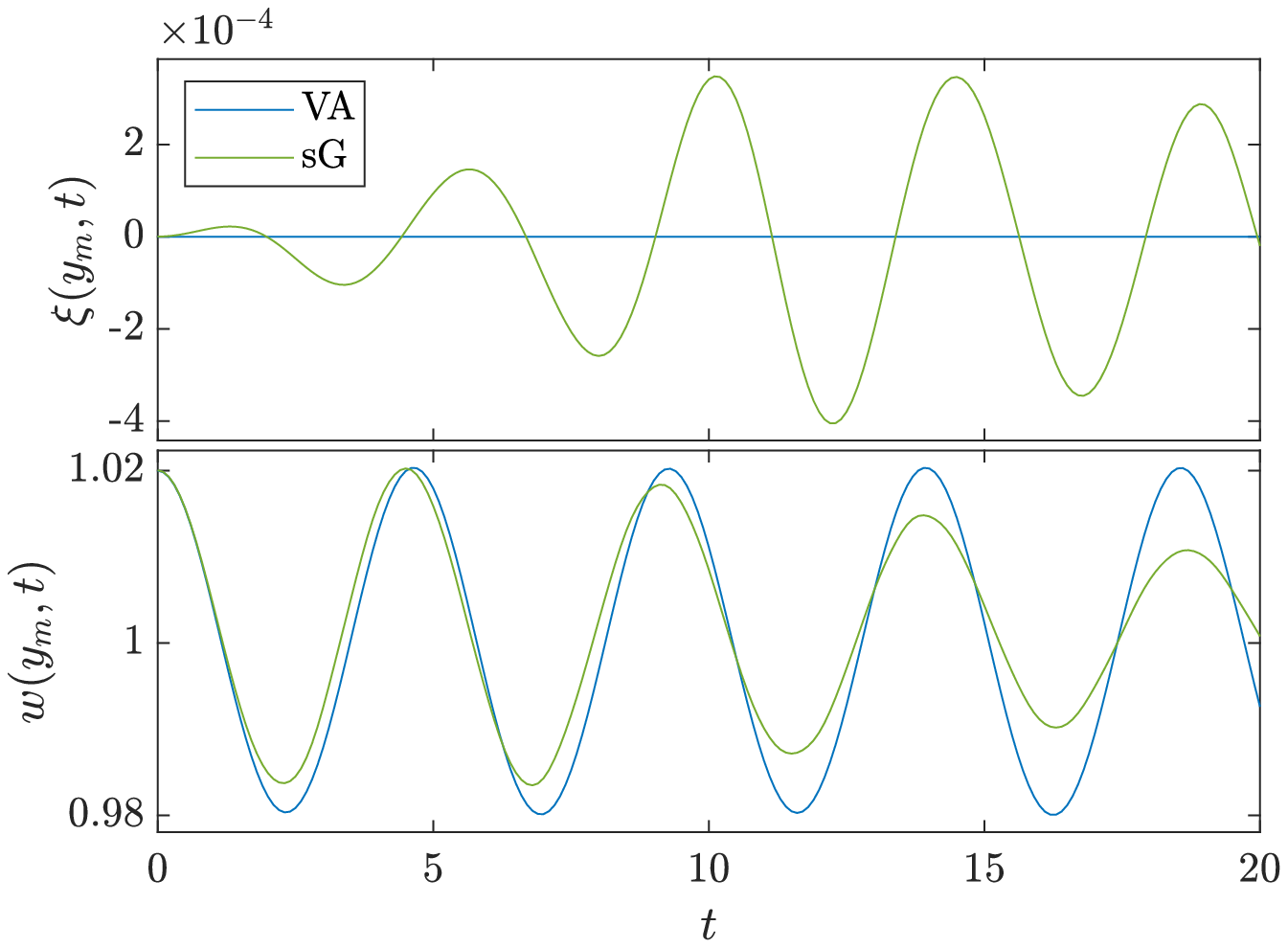}
~
\caption{(Color online)
Similar to Fig.~\ref{fig:VAAISG2} but for a perturbation on the (inverse)
width only corresponding to an initial condition of the form
$w(y,0)=\varepsilon \sin(n\pi y/(2\ell))$ and $\xi(y,0)=\dot\xi(y,0)=\dot w(y,0)=0$ 
with $n=10$, $\ell=20$, and $\varepsilon=0.02$.
Only panels and curves corresponding to the full sG Eq.~(\ref{eq:sG1D}) and 
the stripe VA reduction of Eqs.~(\ref{eq8}) and (\ref{eq9}) are shown.
}
\label{fig:VAAISG3}
\end{center}
\end{figure}

We now compare the reduced VA dynamics of Eqs.~(\ref{eq8}) and (\ref{eq9}) with 
the full dynamics of the 2D sG, Eq.~(\ref{eq:2DsG}), for the single kink (or antikink) 
stripe, and also compare the results with the reduced AI dynamics presented
in Ref.~\cite{our_KD_AI}.
Figure~\ref{fig:VAAISG2} depicts the spatiotemporal evolution of the stripe's 
location $\xi(y,t)$ (left top subpanels) and (inverse) width $w(y,t)$ (right 
top subpanels) for the full 2D sG, Eq.~(\ref{eq:2DsG}) (first row of panels),
the effective VA, Eqs.~(\ref{eq8}) and (\ref{eq9}) (second row of panels), and
the AI reduction of Ref.~\cite{our_KD_AI} (third row of panels).
In the latter, only the center transverse dynamics is present in the
form $\xi_{tt}=\sigma \xi_{yy}$.
The extraction of the position and (inverse) width from the full sG PDE dynamics
was obtained by (least-squares) fitting of a stripe with arbitrary location
$\xi(y,t)$ and (inverse) width $w(y,t)$ for all depicted times.
The kink stripe is initiated using a perturbation of the position of the form
$\xi(y,0)=\varepsilon \sin(n\pi y/(2\ell))$, $w(y,0)=1$, and
$\dot\xi(y,0)=\dot w(y,0)=0$
on the domain $(x,y)\in[-\ell,\ell]\times[-\ell,\ell]$, 
with periodic boundary conditions in $y$ and Dirichlet boundary conditions in $x$. 
In particular, for the example shown in Fig.~\ref{fig:VAAISG2}, the 
parameters of the configuration correspond to $n=10$, $\ell=20$, and $\varepsilon=1$. 
We have verified that the results shown in this figure are typical and that, 
for a wide range of initial perturbations, similar results are obtained (results not shown here).

As it can be observed from the figure, both the VA and AI reductions are able
to accurately predict the undulations of the stripe. However, upon closer
inspection, the full dynamics of the stripe undulations shows a slight 
reduction on the undulation's magnitude since some of the solution's energy 
is transferred to oscillations of the (inverse) width, the so-called ``necking''
dynamics. By construction, the AI assumes that the (inverse) width is constant
($w=1$) and, thus, only the kink center undulates at constant amplitude.
On the other hand, our VA approach does allow for a transfer of energy between
center undulations and necking resulting in a more accurate description of the full
undulation-necking dynamics. 
One can clearly discern the relevant behavior in the bottom two panels of 
Fig.~\ref{fig:VAAISG2}, where we depict cuts of the position and (inverse) 
width for fixed $y$ chosen at the first maximum of the initial condition (see vertical 
dashed lines in the top panels). In these panels it is evident that the AI model 
has constant amplitude undulations of $\xi(y,t)$,
while our VA approach is able to more accurately follow the exchange of energy
between undulation and necking. In fact, not only does the VA closely follow the
undulation amplitude of the stripe transverse position, 
but also it captures more accurately its temporal 
oscillation than the AI (note that the AI tends to slightly underestimate the oscillation 
frequency), and it is also able to qualitatively follow the full necking dynamics.

The reduced VA of Eqs.~(\ref{eq8}) and (\ref{eq9}) describes the effective coupling
between stripe center position undulation and necking dynamics. However, it is 
interesting to note that if one starts with a straight stripe with no undulations 
of the stripe center [$\xi(y,0)=$ constant], then the dynamics
results in no energy transfer  
from necking to undulations, since $\xi_y(y,t)$ will remain zero for all times.
We depict precisely this case in Fig.~\ref{fig:VAAISG3}, where the initial condition
corresponds to a straight stripe $\xi(y,0)=0$ and the perturbation is introduced
in the necking in the same way we introduced the undulation perturbation in
Fig.~\ref{fig:VAAISG2}.
As expected, in the absence of initial undulations, the VA evolution shows 
(see the second row of panels) that the necking dynamics does not induce any 
center undulations for later times. However, the full numerics (see top row of panels)
indicates that the full sG PDE does transfer energy from necking to
center undulations even for initially straight stripes. Nonetheless, we note 
that this transfer is very weak (undulations of the order of $10^{-4}$ for 
the times and parameters of the figure).
Furthermore, the results depicted in Fig.~\ref{fig:VAAISG2}, which we also checked
for other parameter values (results not shown here), indicate that the VA is
successful in reproducing the overall necking dynamics and, in particular, the correct 
frequency of the necking oscillations (see bottom panel). 
We also note that the oscillation of the width, even for the case of a 1D 
sG kink, is not part of the point spectrum, but rather
corresponds to a mode at the edge of its continuous spectrum~\cite{IM}. 
Given the relevant frequency, these width oscillations tend to decay (see bottom panel) 
as energy is transferred to undulations and also, more importantly, as it is shed in the form
of radiation of linear dispersive wavepackets to the background. Further examination 
of the decay of the width oscillations for a 1D kink shows a decay law $\propto t^{-1/2}$
(results not shown here) suggesting that the relevant mechanism involves the generation 
of the 2nd frequency harmonic, which resides in the continuous spectrum and leads to 
resonance with a corresponding mode thereof.

It is also relevant to point out 
that it is possible to employ a multiscale perturbation
approach to describe the kink stripe's transverse dynamics. Indeed,  
motivated by the work of Ref.~\cite{pesenson}, we present such a perturbation 
approach in Appendix~\ref{appendix2}. Through multiple scale dynamics 
along the $y$ direction of a quasi-1D kink stripe along the $x$-direction,
this perturbation approach is able to adequately capture traveling perturbations
on top of the (unperturbed) undulations of the kink stripe, as is
detailed below.

\subsection{Stripe interactions}
\label{sec:stripe_interaction}

\begin{figure*}[htb] 
\begin{center}
\includegraphics[width=1.9\columnwidth]{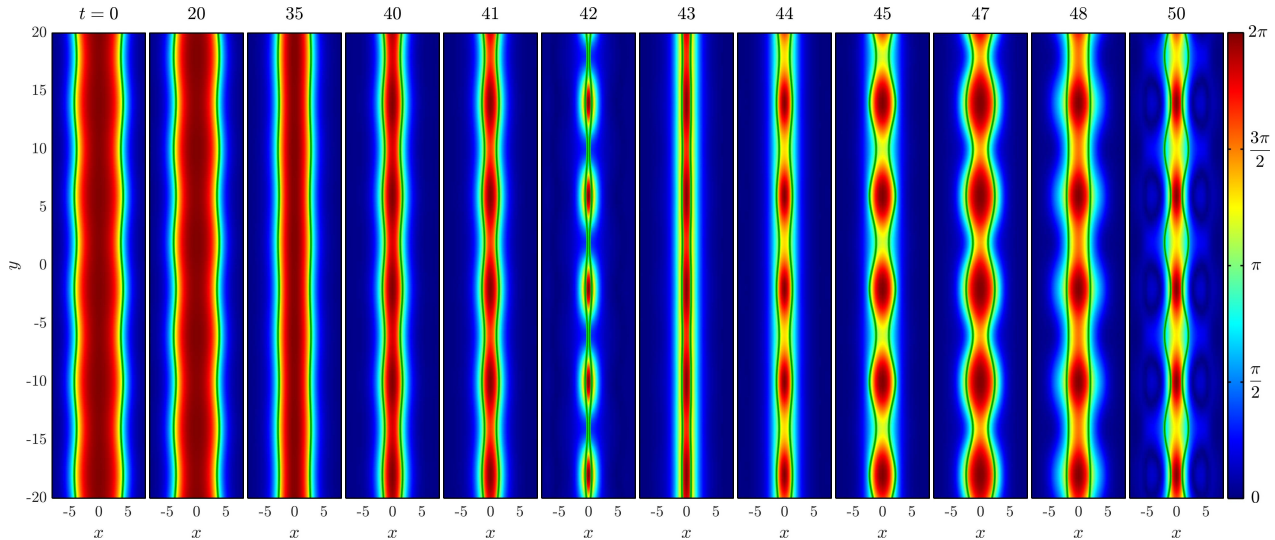}
\caption{(Color online)
Dynamical evolution of two sG stripes. The kink and antikink stripes are
symmetrically perturbed from a straight and parallel configuration such
the stripes have initial positions given by
$-\xi_1(y) = \xi_2(y) = \xi_0 + \varepsilon\sin(2\pi n/\ell)$ for a
separation between the stripes of $2\xi_0=8$ and where the mode
$n=5$ with amplitude $\varepsilon=0.5$ has been excited on the domain
$(x,y)\in[-\ell,\ell]\times[-\ell,\ell]$.
The full sG dynamics of Eq.~(\ref{eq:2DsG}) with $\sigma=1$ is depicted by
the background color at the indicated times.
The corresponding results of the reduced VA model (\ref{eq:VA_2stripes})
are overlaid by the (green) curves.
The stripes collide around $t\approx 42$.
A movie depicting the full stripe dynamics and the reduced VA model
is included in the supplemental material; see movie {\tt 2stripes-mode1.gif}.
}
\label{fig:2stripes_mode1}
\end{center}
\end{figure*}

We now turn to the stripe-stripe interaction case. To this end, we consider 
a 2D ansatz given by extending the symmetric 1D K-AK ansatz of Eq.~(\ref{eqKAK}) 
into 2D:
\begin{equation}
\label{eq:KAK2D}
u(x,y,t)=4\tan^{-1} \left[\sinh(w \xi) \sech(wx)\right], 
\end{equation}
where now the positions and (inverse) widths of the kinks are $(y,t)$-dependent
according to $\xi(y,t)=-\xi_1(y,t)=\xi_2(y,t)$ and $w(y,t)=w_1(y,t)=w_2(y,t)$.
%
\begin{strip}
\noindent
Then, the averaging of Lagrangian (\ref{eq:LagLy}) on this ansatz yields a variant of the 
Lagrangian~(\ref{Lg1D}) of the form:
\begin{equation}
\label{eq11}
L_{2D}=\frac{1}{2}\left(w_t^2-\sigma w_y^2\right)N+\frac{1}{2}\left(\xi_{t}^2
-\sigma \xi_{y}^2\right)M 
+\left(w_t\xi_{t}-\sigma w_y \xi_{y}\right)I-{\cal V},
\end{equation}
where the terms $N$, $I$, $M$, and ${\cal V}$ are given, respectively, as in 
Eqs.~(\ref{eq:NIMV1})--(\ref{eq:NIMV4}), but with the important distinction that
now $\xi(y,t)$ and $w(y,t)$ also depend on the $y$-coordinate.
%
%
Taking full variations of the 2D Lagrangian~(\ref{eq11}) in terms of $\xi$ and $w$ yields:
\begin{eqnarray}
M\xi_{tt}+Iw_{tt}=\sigma M\xi_{yy}+\sigma Iw_{yy}-\frac{1}{2}M_{\xi}(\xi_{t}^2-\sigma \xi_{y}^2)
+(w_t^2-\sigma w_y^2)\left(\frac{1}{2}N_{\xi}-I_w\right)-M_w(\xi_{t}w_t
-\sigma \xi_{y}w_y) -{\cal V}_{\xi}, 
\label{eq:KAKxiw1}
\\
\label{eq:KAKxiw2}
I\xi_{tt}+Nw_{tt}=\sigma Nw_{yy}+\sigma I\xi_{yy}-\frac{1}{2}N_w(w_t^2-\sigma w_y^2)
+(\xi_{t}^2-\sigma \xi_{y}^2)\left(\frac{1}{2}M_w-I_{\xi}\right)-N_{\xi}(\xi_{t}w_t
-\sigma \xi_{y}w_y)-{\cal V}_w. 
\end{eqnarray}
%
\end{strip}
\noindent
Equations~(\ref{eq:KAKxiw1}) and (\ref{eq:KAKxiw2}) describe the reduced
(undulation and necking) dynamics for the K-AK stripe interactions in the 
symmetric $\xi(y,t)=-\xi_1(y,t)=\xi_2(y,t)$ case.

\begin{figure*}[t] 
\begin{center}
\includegraphics[width=1.9\columnwidth]{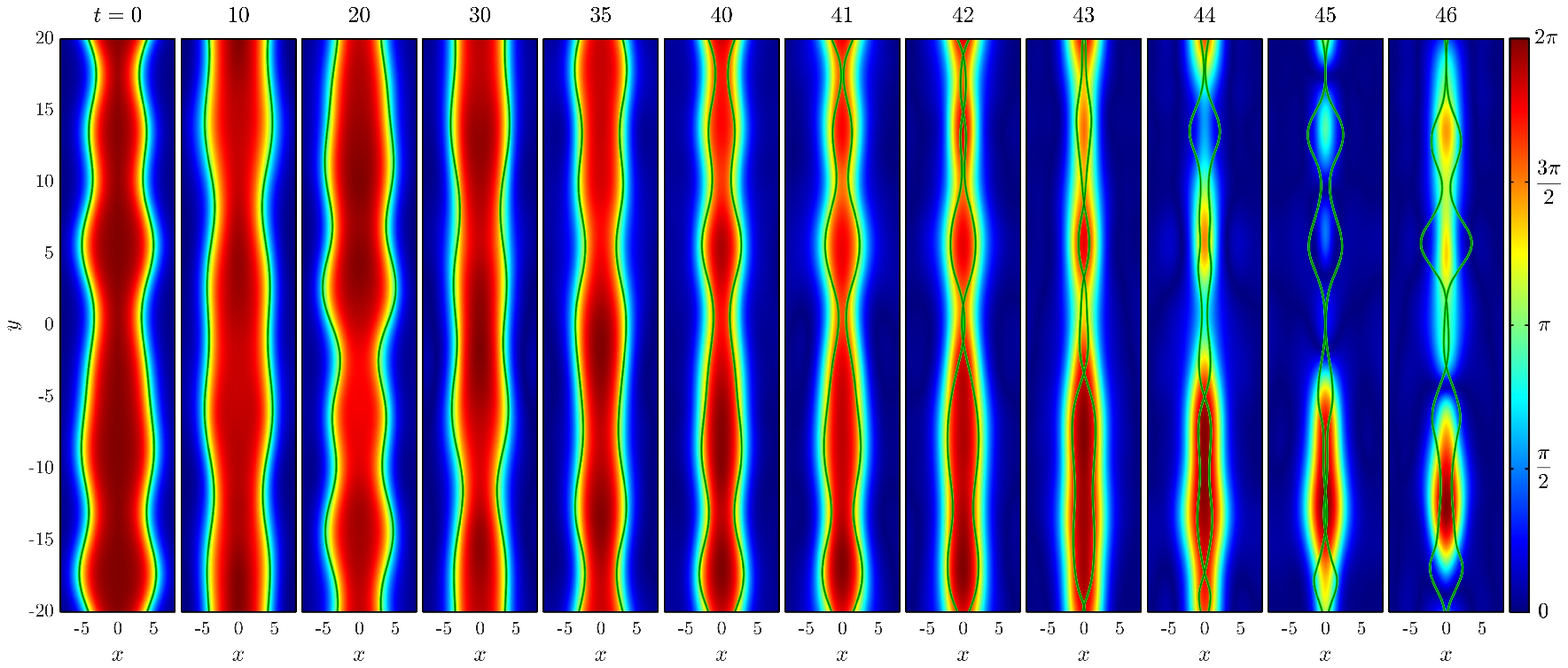}
\caption{(Color online)
Similar to Fig.~\ref{fig:2stripes_mode1} but for a perturbation containing
the first five modes.
A movie depicting the full stripe dynamics and the reduced VA model
is included in the supplemental material; see movie {\tt 2stripes-mode5.gif}.
}
\label{fig:2stripes_mode5}
\end{center}
\end{figure*}

Similar to the simplification that we performed in the case of the
1D kink to obtain a Newtonian-type equation for the kink center, 
assuming that $w=1$ in the Lagrangian~(\ref{eq11}) yields the following
reduced equation of motion for the K-AK stripe position:
\begin{equation}
\label{eq:VA_2stripes}
M\xi_{tt}=\sigma M\xi_{yy}-\frac{1}{2}M_{\xi}(\xi_{t}^2-\sigma \xi_{y}^2)-{\cal V}_{\xi}.
\end{equation}
It is important to note that, since the K-AK complex is driven by mutual {\em attractive}
interactions, the K-AK stripe will evolve towards the (local) collision of the stripes 
close to the location of the narrower distance between the stripes. Thus, by construction, the
dynamics that can be followed by our VA methodology will be limited to times before
any portion of the stripes get into close contact (collision).
Therefore, if only considering stripe undulations, it is not necessary to employ the 
full coupled system (\ref{eq:KAKxiw1}) and (\ref{eq:KAKxiw2}) since the small
variations of the width will have only a weak effect on the short-term dynamics
before collision.

We now compare the full dynamics of the K-AK interactions with the simplified
dynamical reduction (\ref{eq:VA_2stripes}).
Figure~\ref{fig:2stripes_mode1} depicts the evolution of a K-AK stripe
combination whose initial condition corresponds to two straight and
parallel stripes that have been perturbed by a single spatial mode. 
As the figure strongly suggests, the reduced VA model (\ref{eq:VA_2stripes}) is able 
to capture, simultaneously, the periodic (in time) undulations of the stripes and
their mutual attraction.
It is interesting that, in this case, the reduced VA model is able to capture
some of the dynamics close to and just after the collision ($41<t<47$).
To further evidence the value of the reduced VA model in capturing the full
dynamics, we depict in Fig.~\ref{fig:2stripes_mode5} a similar case
to the one in Fig.~\ref{fig:2stripes_mode1}, but where the
first five spatial modes have been included in the initial perturbation.
Again, the reduced VA model captures remarkably well both the undulations
and the interaction dynamics of the stripes until shortly after they first 
collide around $t\approx 41$ for $y\approx 17.5$.

\section{Conclusions \& Future Work}
\label{sec:conclu}

Through a variational approximation methodology, we have studied the 
dynamics of sine-Gordon kink and antikink interactions in the
one- and two-dimensional sine-Gordon equations. This dynamical reduction, based
on following the time-dependent widths and locations of the kink and antikink,
leads to an effective attractive interaction potential
between the kink and the antikink. The resulting Newtonian-type dynamical
system was qualitatively as well as quantitatively used
to accurately describe the phase- and real-space dynamics of the kink-antikink
separation where a separatrix demarks the boundary between bounded 
(periodic) and unbounded kink-antikink bounces in the one-dimensional realm.

We extended this variational approximation type of consideration
to the case of two-dimensional
kink stripes where their width and positions are allowed to not only vary 
in time but also in space (i.e., along the transverse direction $y$).
We find remarkable agreement between the full dynamics of the original
sine-Gordon PDE and the reduced stripe equations of motion. Indeed,
this was found to significantly improve the earlier
adiabatic-invariant
perspective of Ref.~\cite{our_KD_AI}, which only examined the transverse
dynamics of the kink center position. 
The two-dimensional analysis is carried out for the elliptic and hyperbolic
Laplacian cases showing that kink stripes are always unstable in the
hyperbolic case, while generically stable in the elliptic case.
The two fundamental ingredients, namely the understanding of K-AK
interactions and that of the single-stripe transverse dynamics
were subsequently brought together confirming that the variational
analysis is able to capture the complex interplay between the two,
ultimately leading through the interactions to the formation of
localized blobs somewhat reminiscent of the radial kinks
of Ref.~\cite{caputo,our_KD_AI} (see also the discussion and references therein). 

There is a number of potential avenues for further research related to the present work.
For instance, it would be interesting to check whether the intrinsic instability of
the kink stripes in the hyperbolic dispersion regime could be tamed by suitable
introduction of dissipative terms~\cite{kivsharmalomed,josephson_dissip}. 
It would also be interesting to study the long-time dynamics of stripe undulations, 
as our results suggest that the coupling between undulations and necking 
results in a small amount of radiation loss towards the background inducing
a weak damping of the undulation oscillations over time. Understanding 
also the long-term evolution of the stripes and the asymptotic fate
of the system may have not been a focal point herein (as it was
outside the reach of the present theory), yet it constitutes an interesting
question for further study. 
Extending the considerations herein to other
settings, such as ones involving radial kinks~\cite{our_KD_AI} or the
so-called pulsons~\cite{malomed_pulsons}, superluminally moving (``tachyonic'') 
kinks~\cite{malomed_PD_1991} or multi-kinks could
be a separate theme of study. 
Moreover, based on the emerging
understanding of both kink stripes (per the work herein) and
2D-breathers (see, e.g., Ref.~\cite{malo}), considerations of interactions
between these structures (see, e.g., the relevant studies~\cite{neshev1,neshev2} 
in the context of the nonlinear Schr{\"o}dinger model), and
with other structures, such as impurity modes~\cite{kivsh}, would
constitute yet another interesting direction of future study.

\appendix

\section{An alternative approach to K-AK identification}
\label{appendix1}

In Sec.~\ref{sec:KAK1DVA} we used Eq.~(\ref{eq:x0FromAnsatz}) 
to track the K-AK half separation distance or, equivalently, the 
AK position. Another way to track the antikink (or kink) position 
is to use the right (or left) inflection point of the  K-AK solution 
[cf.~Eq.~(\ref{eq:KAK_exact})] or breather solution [Eq.~(\ref{eq:1D_breather})];  
see Appendix A of Ref.~\cite{nlsns} for a discussion of the use of the inflection 
point as a measure of the kink center for an asymmetric kink in a $\phi^8$ theory.
One benefit of using the inflection point as the location of a kink is
that it does not depend on a particular ansatz. It can be tracked for
any two-soliton interaction, using either an exact formula when
available or numerically for arbitrary models featuring K-AK
interactions. Finally, it can be tracked through the entire evolution
of the relevant interaction, unlike when the kink location  is potentially
calculated as the intersection of the kink with a particular value of $u$.
%
Naturally, as the K and AK approach each other, one cannot truly
consider them as having separate identities; still this method of
tracking kink location can be applied successfully to many different
models and can provide some insight into the relevant phenomenology.

We start with the K-AK solution of Eq.~(\ref{eq:KAK_exact}). Setting
$u_{xx}=0$, and solving for $x$ to get the rightmost inflection point
of the exact solution, we obtain:
\begin{equation}
x(t)= \sqrt{1-v^2}\,\ln\left(\sqrt{\frac{2\sigma_1 +2s_1+3v^2 }{v^2 }}\right),
\label{AKsolution}
\end{equation}
where
\begin{eqnarray}
\notag
s_1 &=& \sqrt{{\left(v^2 +\sigma_1 \right)}\,{\left(2\,v^2 +\sigma_1 \right)}},
\notag
\\
\notag
\sigma_1 &=&\sinh^2\left(\frac{t\,v}{\sqrt{1-v^2 }}\right).
\end{eqnarray}
Similarly, for the breather solution of Eq.~(\ref{eq:1D_breather}), we get:
\begin{equation}
x(t)= \frac{1}{\sqrt{1-\omega^2}}
\ln \left(\sqrt{\frac{2\sigma_2 +2s_2-2\omega^2 \,\sigma_2 +3\omega^2 }{\omega^2 }}\right),~~~~~~~
\label{rightSolitonPosition}
\end{equation}
where
\begin{eqnarray}
s_2 &=& \sqrt{{\left(-\omega^2 \,\sigma_2 +\omega^2 +\sigma_2 \right)}\,{\left(-\omega^2 \,\sigma_2 
+2\,\omega^2 +\sigma_2 \right)}},
\notag
\\
\notag
\sigma_2 &=&{\cos^2 \left(\omega t\right)}.
\end{eqnarray}
For both cases, the antikink is to the right of the kink for $u(x,t)>0$, and vice-versa for 
$u(x,t)<0$. When $u(x,t)$ goes from positive to negative, we can identify the incoming 
antikink with the outgoing kink; we will call this the \textit{right soliton}, and similarly 
the incoming kink and outgoing antikink the \textit{left soliton}. Thus, $x(t)$ in 
Eq.~(\ref{AKsolution}) or Eq.~(\ref{rightSolitonPosition}) represents the position of the right 
soliton, and $-x(t)$ represents the position of the left soliton. 

\begin{figure}[t] 
\begin{center}
\hskip -0.6cm
\includegraphics[width=4.5cm,height=3.8cm]{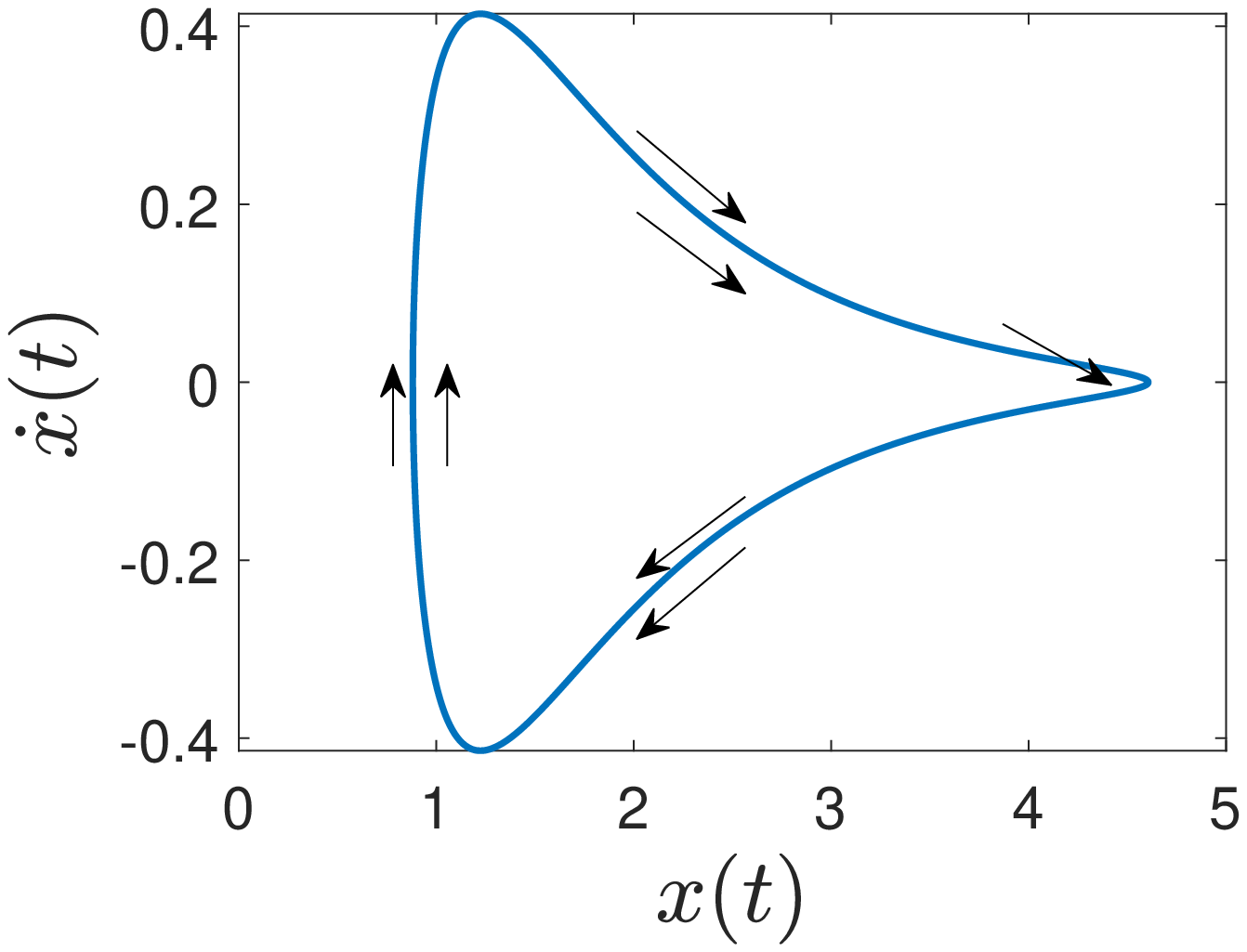}
\hskip -0.2cm
\includegraphics[width=4.5cm,height=3.8cm]{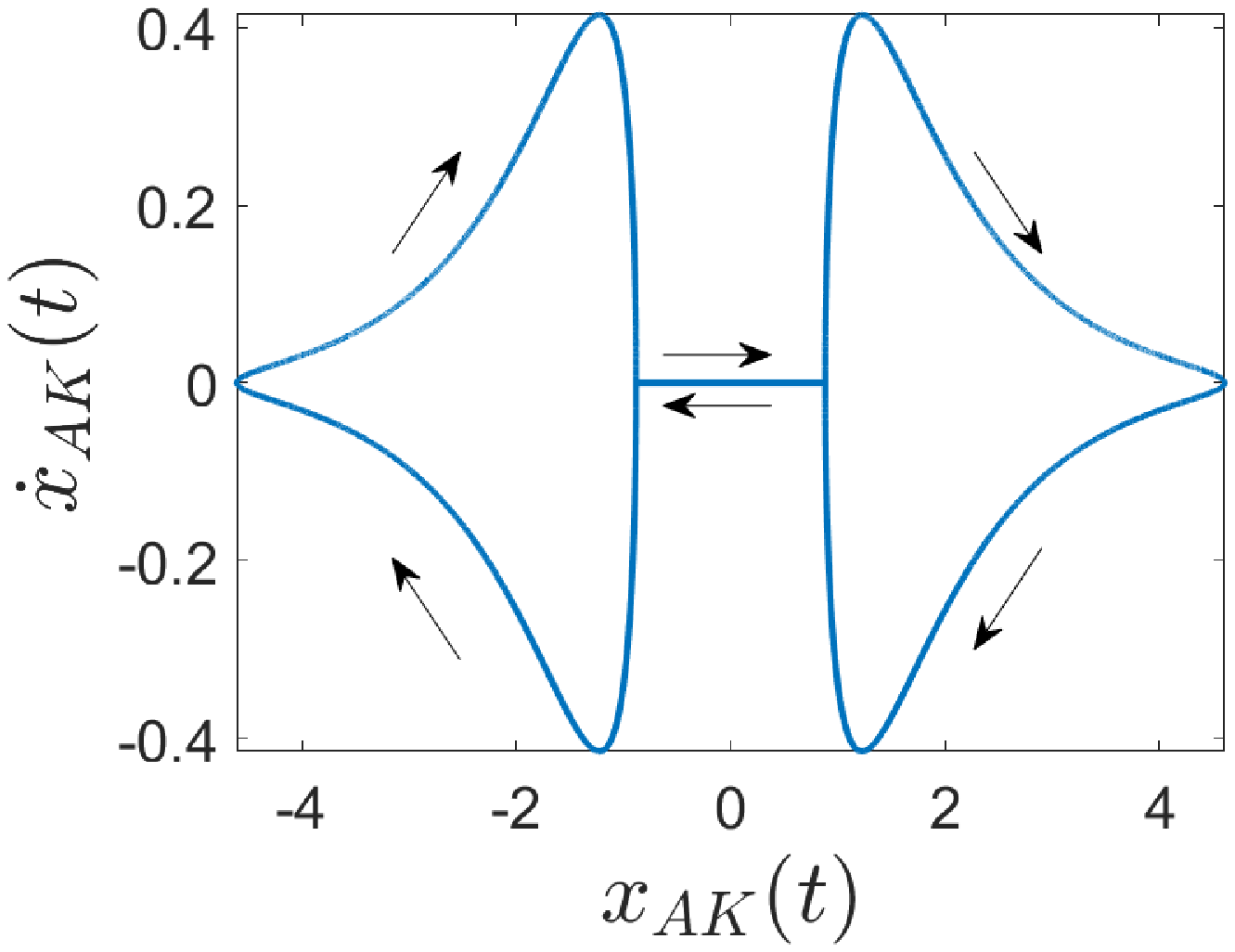}
\hskip -0.9cm
\null
\caption{Two perspectives of the phase plot of a breather cycle for
  $\omega=0.02$, using the inflection point for kink/antikink location. The
  phase plot on the left uses $x(t)$ from
  Eq.~(\ref{rightSolitonPosition}) and so tracks the right
  soliton. The arrows indicate that one breather cycle corresponds to
  twice around the loop (arrows outside for the first time around the
  loop, arrows inside for the second time around the loop). The plot on the right uses $x_{AK}(t)$, the antikink position. There is an instantaneous jump in the antikink position at the points where the position reaches its closest approach to zero (and the velocity reaches zero).}
\label{fig:twoWaysPhase}
\end{center}
\end{figure}

Alternatively, using the inflection point for the soliton position in order to create a phase plot that 
is comparable to the ones in Fig.~\ref{fig:PhaseSpace}, we need to identify the incoming antikink 
with the outgoing antikink. To track the antikink we need to modify Eqs.~(\ref{AKsolution}) and 
(\ref{rightSolitonPosition}) as follows. We define the antikink position $x_{AK}(t)$ to be $x(t)$ 
when $u(x,t)>0$, and $-x(t)$ 
when $u(x,t)<0$. In Fig.~\ref{fig:twoWaysPhase} we 
show two perspectives of a phase plot of a single breather trajectory using $\omega=0.02$.
The left panel of Fig.~\ref{fig:twoWaysPhase} shows a phase plot using
$x(t)$ from Eq.~(\ref{rightSolitonPosition}) to track the right
soliton. What we see is that the $x$-coordinate of the right soliton
never reaches zero (and similarly for the left soliton). This leads to
an interpretation that the right and left solitons reach a minimum
(non zero) separation distance and then ``bounce'' off of each other.
%
However, notice that the attractive interaction picture 
discussed in Sec.~\ref{sec:KAK1DVA} is expected to fail when the distance 
between the solitons becomes so small that the solitons start to overlap 
(in such a case the particle picture becomes irrelevant, as the two solitons 
cannot be considered as individual Newtonian particles). 
The right panel shows a phase plot using the antikink position
$x_{AK}(t)$. In this case there is an instantaneous jump in position
from positive to negative when the antikink position reaches its
closest approach to zero (and the antikink velocity reaches
zero). Since this instantaneous jump appears to be unphysical, the first
perspective seems to make more sense when using the inflection point
to measure the soliton location, and so we will use the type of phase plot in the left panel of 
Fig.~\ref{fig:twoWaysPhase} for most of the rest of this subsection.

\begin{figure}
\begin{center}
\includegraphics[width=0.75\columnwidth]{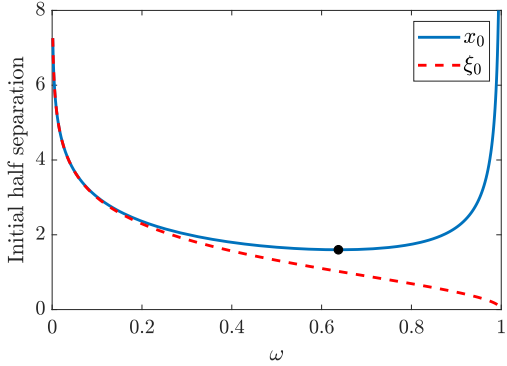}
\vspace{-0.3cm}
\end{center} 
\caption{Initial half separation distance as a function of the breather parameter $\omega$ for the 
variational approach ($\xi_0$, red, dashed) and for the inflection point approach ($x_0$, blue, 
solid), using Eqs.~(\ref{x0versusW}) and (\ref{xiVersusW}).
Contrary to the variational approach, the inflection point approach displays a
minimum (shown by a black dot).} 
\label{x0versusM}
\end{figure}

For the breather case, the initial half-separation $x_0=x(0)$ (inflection point method), which is 
also the maximum half-separation, occurs when $t=0$, so from Eq.~(\ref{rightSolitonPosition}) 
we get:
\begin{equation}
\label{x0versusW}
x_0=\frac{1}{\sqrt{1-\omega^2}}\ln\left(\sqrt{\frac{2\,\sqrt{\omega^2 +1}
+\omega^2 +2}{\omega^2 }}\right),
\end{equation}
which gives us a relationship between $x_0$ and the parameter $\omega$. 
We can also find the initial half separation distance $\xi_0$ as a function of the 
parameter $\omega$ for the variational approach using Eq.~(\ref{eq:x0FromAnsatz}), 
with $t=0$ and $u(x,t)$ the breather solution from Eq.~(\ref{eq:1D_breather}). The result is:
\begin{equation}
\xi_0=\sinh^{-1}\left(\frac{\sqrt{1-\omega^2 }}{\omega}\right).
\label{xiVersusW}
\end{equation}
In Fig.~\ref{x0versusM} we show graphs of the initial half
separation distance as a function of $\omega$, for both the variational
approach and the inflection point method. For both, the value $\omega=0.1$
gives rise to a half separation of approximately $3$, and the two
curves are nearly identical for $\omega<0.1$ and a half separation greater
than $3$.

\begin{figure}
\begin{center}
\includegraphics[width=0.75\columnwidth]{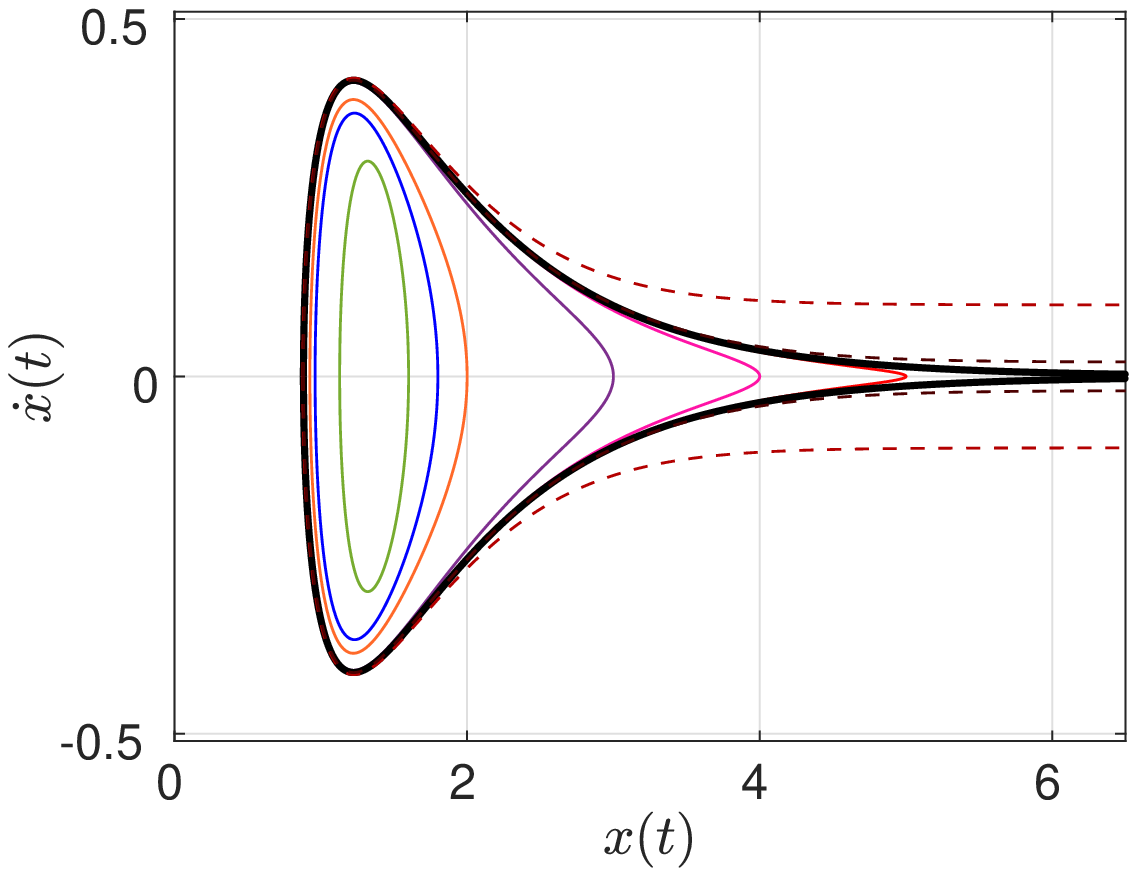}
\includegraphics[width=0.75\columnwidth]{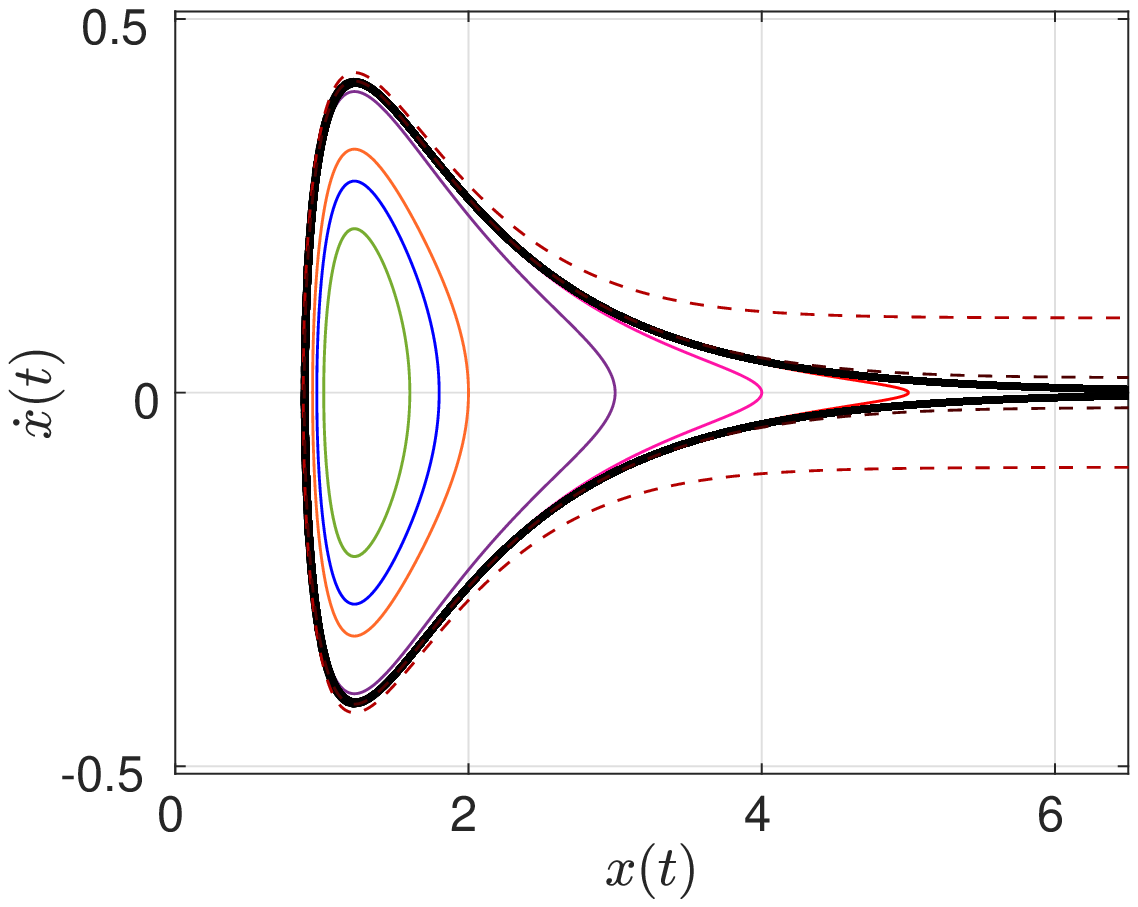}
\end{center} 
\caption{(Color online)
Phase portraits 
for breather (solid lines) and K-AK (dashed lines) solutions
from PDE (top) and from fitted ODE (bottom). Top panel uses the right soliton location 
of the PDE and the bottom panel uses the ODE acceleration law of Eq.~(\ref{mainODE}). 
The outer loops of the PDE are accurately reproduced by the ODE and the inner loops 
are qualitatively similar. The thicker black lines in both represent the separatrix.}
\label{fittedPhase}
\end{figure}

For the inflection point method there is a minimum value for $x_0$ as a function of $\omega$;
the minimum value is $x_{0,\rm min}=1.5994$ and it occurs at $\omega_{\rm min}=0.63739$, 
as found by differentiating Eq.~(\ref{x0versusW}); see the black dot in Fig.~\ref{x0versusM}.
One implication of this minimum is that none of the solutions contained in the family of solutions 
given by Eq.~(\ref{eq:1D_breather}) corresponds to a $x_0$ such that $x_0<x_{0,\rm min}$. 
In Fig.~\ref{fittedPhase} (top panel) we show a number of $x(t)$ trajectories, creating a phase 
portrait for the position of the right soliton. All of the breather trajectories have been created 
using $\omega$ values for which $\omega<\omega_{\rm min}$. 
There are also three K-AK trajectories with 
velocities $0\le v\le 0.1$, where, similar to the single kink, $v$ is the initial K-AK
velocity corresponding to an infinitely large separation $x=+\infty$ (dashed curves), 
with the smallest being very close to zero and hence representing the 
separatrix (thick black line).

\begin{figure}
\begin{center}
\includegraphics[width=0.85\columnwidth]{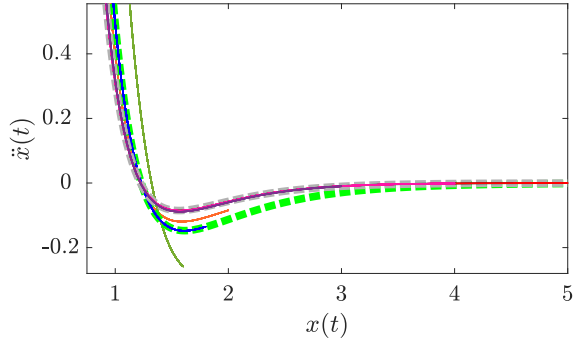}
\end{center} 
\caption{(Color online)
Acceleration versus position (right soliton) for different
initial half-separation values $x_0$ (breather case) and $v$ values (K-AK case). 
The colors of the different orbits correspond to the same
orbits and colors in Fig.~\ref{fittedPhase}.
Breather curves with $x_0=3, 4, 5$ and the three K-AK 
curves with $0\le v\le 0.1$ coincide at the scale shown. 
The three fore-shortened thin solid curves at the bottom correspond 
to the smallest three separation values of 
$x_0=1.6$ (dark green), $x_0=1.8$ (blue), and $x_0=2.0$ (red).
The upper (gray) thick dashed line is fit to case $x_0=5$ corresponding to
Eq.~(\ref{mainODE}) and the lower (light green) thick dashed line is fit to case $x_0=1.8$
corresponding to Eq.~(\ref{mainODE2}).
}
\label{accelVsX_x0_all}
\end{figure}

Note that trajectories in Fig.~\ref{fittedPhase} (top panel)  do not
intersect, partially justifying the interpretation of the soliton
motion as particle motion. However, the smallest inner (green) loop uses
$x_0=x_{0,\rm min}$ and so no trajectory is possible inside this smallest
loop. If this were the phase portrait for a particle we would expect
more nested loops inside this one, and a fixed point inside of these
loops, but no such fixed point exists for the position of the right
soliton (such a fixed point would be represented in the limit by a
breather, for which the inflection point is practically moving  vertically
in the $(u,x)$ plane). Indeed, while the inflection point methodology has some
notable advantages given its widespread potential numerical
applicability, these issues, i.e., the existence of such a fictitious fixed point
---as is also discussed below--- and the
inability to capture the entire phase space of the kink motion somewhat
limit its applicability and require perhaps additional future consideration
towards its broader application.

Finally, we would like to note that when using $\omega$ values for which 
$\omega>\omega_{\rm min}$, the system experiences small-amplitude breathers 
which have increasingly larger spatial extent. This creates multiple small 
loops which overlap in the ``phase portrait'' (results not shown here). 
Hence these values of $\omega$ do not result in behavior which could be 
interpreted as particle motion. 
%

In order to derive an ODE that replicates the behavior of the right (or left) soliton, we need 
to determine the force (or acceleration) for each value of the separation distance $x(t)$. 
We could use the second derivative of Eq.~(\ref{AKsolution}) for the K-AK 
and Eq.~(\ref{rightSolitonPosition}) for the breather solutions shown in the portrait 
in Fig.~\ref{fittedPhase} (top panel), and plot $x''(t)$ against 
$x(t)$ to get a possible force/acceleration law. 
However, different loops give different results; for initial breather separations 
of $3$ or more (outer breather loops in the phase portrait) and for K-AK solutions 
with $0\le v\le 0.2$ (dashed lines in the phase portrait),
the acceleration versus position graphs are 
nearly indistinguishable, but for the inner breather loops the results start to differ. 
We show the resulting acceleration-position curves in Fig.~\ref{accelVsX_x0_all} 
---the colors correspond to the colors of the loops in Fig.~\ref{fittedPhase}, 
top panel. The three solid curves with lowest minima (red, blue, and dark green)
correspond to the three inner loops in the top panel of Fig.~\ref{fittedPhase}, 
for which $x_0<3$.

Because the force (acceleration) law is not consistent we must
conclude that, within the present approach, no force law (and hence 
no ODE) will completely replicate the entire phase portrait of soliton position created from the PDE 
itself. However, for initial breather separations of $3$ or more and for K-AK solutions with 
$0\le v\le0.2$, the force law is nearly the same, and so in this regime (the outer breather loops 
and first few K-AK solutions of the phase portrait) a reasonably accurate two degree of freedom 
``particle'' description of the soliton position is possible. 

For the breather case with $x_0\ge3$ and the K-AK case with $0\le v\le 0.2$ we can get an approximate (though quite accurate) analytic acceleration expression by fitting a curve of the form $a\exp(-b x)-c\exp(-d x)$. This form is suggested by appropriate logarithmic plots (not shown here) that display straight line behavior for acceleration versus position when $x$ is either large ($x\ge3$) or small ($x\le1$).
%
In Fig.~\ref{accelVsX_x0_all} we show that we get an excellent fit to
the acceleration versus position plots, either for the base model
[lower (light green) thick dashed line is fit to the case $x_0=5$] or for possible
extensions of the inner loop models [upper (gray) thick dashed line is fit to the
case $x_0=1.8$]. 
The result of fitting the case $x_0=5$ (but also appropriately for any 
$x_0>3$ for breather solutions and for the K-AK cases with $0 \leq v \leq 0.2$) 
yields the ODE model
\begin{equation}
x''=-2.666\exp(-1.848x)+57.05\exp(-4.355x);
\label{mainODE}
\end{equation}
see upper thick dashed (gray) curve in Fig.~\ref{accelVsX_x0_all}.
Fitting a curve to the acceleration data corresponding to $x_0=1.8$ results in the ODE model
\begin{equation}
x''=-2.357\exp(-1.457x)+72.49\exp(-4.2431x);
\label{mainODE2}
\end{equation}
see lower thick dashed (light green) curve in Fig.~\ref{accelVsX_x0_all}.
It should be mentioned that these expressions offer a rather distinct
picture than the well-known asymptotic results for the K-AK
interaction, such as those, e.g., obtained via the so-called Manton
method~\cite{Manton}, according to which:
\begin{equation}
  x''=-4 \exp(-2 x).
  \label{mant}
\end{equation}
While the interaction is attractive at longer range, clearly the use
of the inflection point seems to introduce a distinct (secondary) form
of effective short-range repulsion that has been absent in our
earlier considerations, e.g., in Fig.~\ref{fig:PhaseSpace}.

\begin{figure}[t] 
\begin{center}
\includegraphics[width=0.95\columnwidth]{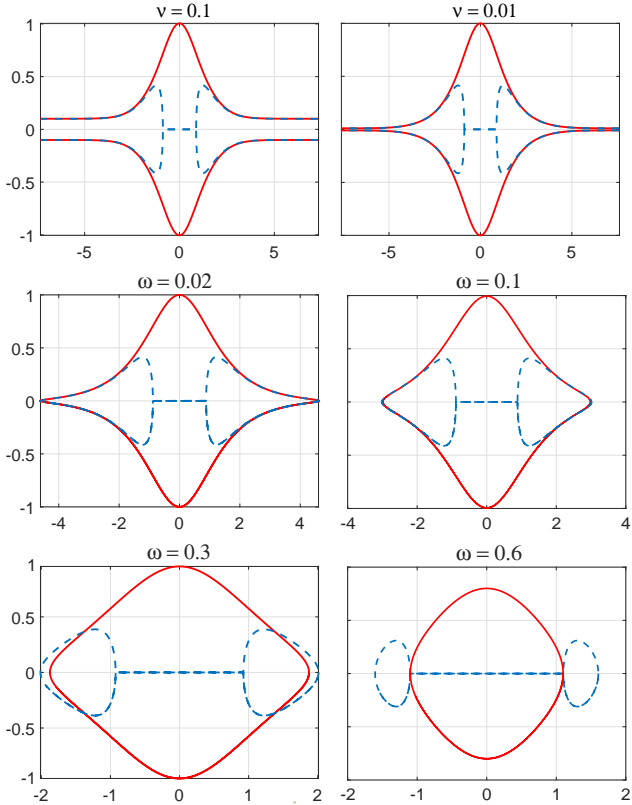}
\end{center} 
\caption{Phase plot comparison of variational ($\xi(t)$, solid red)
  and inflection point ($x_{AK}(t)$, dashed blue) methods of
  calculating the antikink position, using the K-AK and breather sG
  solution in Eqs.~(\ref{eq:KAK_exact}) and
  (\ref{eq:1D_breather}).
  K-AK solutions corresponding to the
  parameter values $v=0.1$, and $0.01$ are shown (top two panels) and
  breather solutions corresponding to the parameter values
  $\omega=0.02, 0.1, 0.3$, and $0.6$ are shown in the bottom four panels.
  In all cases,
  the position is on the horizontal axis and the velocity on the
  vertical axis.
  The agreement is good for antikink
  positions greater than $2$ (in absolute value).}
\label{compareInflVar}
\end{figure}

Nevertheless, the above fitted analytic expressions can be used to
create phase portraits that can be compared to the portrait created
directly from the PDE. In Fig.~\ref{fittedPhase} (bottom panel) we do
so for the case corresponding to Eq.~(\ref{mainODE}).
As can be seen from Fig.~\ref{fittedPhase}, the ODE phase portrait fit to the case $x_0=5$ reproduces the PDE outer loops accurately. 
Though not shown, the ODE phase portrait fit to the case $x_0=1.8$ given in 
Eq.~(\ref{mainODE2}) does a better job of reproducing the PDE inner loops.

For a comparison of the phase portraits for the variational approach
and the inflection point method see Fig.~\ref{compareInflVar} where we
show phase portraits based on the exact sG solutions in
Eq.~(\ref{eq:KAK_exact}) and Eq.~(\ref{eq:1D_breather}). Here we use
the method of the right panel of Fig.~\ref{fig:twoWaysPhase} to show
the phase plots for the inflection point method. The top two panels
show the K-AK cases for $v=0.1, 0.01$, and the middle and bottom
panels show the breather cases for $\omega=0.02, 0.1, 0.3, 0.6$. In all
cases the variational antikink position $\xi(t)$ is depicted in red and the
inflection point antikink position $x_{AK}(t)$ is depicted in blue. We see
that in all cases the agreement between the two is quite good for
antikink position values greater than about $2$ (in absolute value).
Once again, this represents the case where the separation of the kinks
is larger than their individual widths.
The last (bottom right) panel shows the case where $\omega$ is very near
the point where the initial separation is at a minimum using the
inflection point method ---see Fig.~\ref{x0versusM}. For larger $\omega$ the
left and right loops (inflection point method) drift further away from
the origin whereas the variational loops continue to become smaller
approaching the origin. Clearly the two approaches yield fundamentally
distinct results in this limit and it is here where the principal
concerns
raised above regarding the inflection point method are most acute.
It is relevant to add at this point that Fig.~\ref{accelVsX_x0_all}
also predicts exactly in this spatial range the existence of a fixed
point balancing the attraction at larger distances (negative acceleration)
and the repulsion at shorter distances (positive acceleration). It is
fair to say that we are not aware of the presence of such a fixed
point at the PDE level.

In summary, we have discussed in this Appendix
the potential usefulness and robustness of using the inflection point
(and similar PDE diagnostics) at large distances, i.e., ones larger
than the sum of the widths of the colliding kinks. We have also, however,
noted for completeness (and also to alert the potential numerical
practitioners using such PDE models) the issues and caveats that one
needs to pay attention to when using such diagnostics at short
distances, as they may suggest features that may not be identifiable
at the PDE level.


\section{Multiple scales approach to transverse kink dynamics}
\label{appendix_multiscales}

Here we demonstrate how Eqs.~(\ref{eq8}) and (\ref{eq9}) may be 
obtained from a classical multiple scales analysis of Eq.~(\ref{eq:2DsG}). 
In order to do so, one must adopt a geometric optics scaling 
(cf.~the book by Anile et al.~\cite{anile}), 
 which is an approach frequently used to analyze the transverse stability of solitary waves 
(see Refs.~\cite{ostrov-shrira,Infeld-Rowlands,kivshar-pelin} for some early and key examples of this).
To utilize this, we write the phase of the front $g$ given in Eq.~(\ref{eq5}) as
\[
\begin{gathered}
g = w(Y,T)\bigg(x-\frac{\xi(Y,T)}{\varepsilon} \bigg) \equiv w\theta\,, \\[3mm]
{\rm where} \quad  (Y,T) = \varepsilon (y,t)\,, \quad \& \quad \varepsilon \ll 1\,.
\end{gathered}
\]
This is then utilized to construct the guess at a perturbed solution of the form
\[
u = u_0(\theta;w)+\varepsilon u_1(\theta;w)+\varepsilon^2 u_2(\theta;w)+\mathcal{O}(\varepsilon^3)\,,
\]
where the function $u_0$ is defined through Eqs.~(\ref{eq4}) and (\ref{eq5}) with $s = 1$
taken (as polarity does not affect the resulting dynamics)
and the functions $u_i$ represent corrections to this solution.
This leads to the SG Eq.~(\ref{eq:2DsG}) becoming a system
of equations at the various orders $\mathcal{O}(\varepsilon^i)$:
\begin{align}
\mathcal{O}(\varepsilon^0): \quad\, 0 = & \ w^2(1-\xi_T^2-\sigma\xi_Y^2)(u_0)_{\theta \theta}-\sin(u_0)\,,\label{zero-order}\\[5mm]
\mathcal{O}(\varepsilon^1): ~ {\bf L}u_1 =& \ \big( (w\xi_T)_T-\sigma(w\xi_Y)_Y\big)\frac{(u_0)_\theta}{w} \hfill \notag\\[2mm]
&+2(w_T\xi_T-\sigma w_Y\xi_Y)\,, \hfill\label{first-order}
\\[5mm]
\mathcal{O}(\varepsilon^2): ~ {\bf L}u_2 =& \ (w_T\xi_T-\sigma w_Y\xi_Y)\big(2\theta(u_1)_{\theta \theta}+(u_1)_\theta\big)\notag\\[2mm]
&+\big((w\xi_T)_T-\sigma(w\xi_Y)_Y\big)(u_1)_\theta\notag\\[2mm]
&+2\bigg(\frac{w_T\theta}{w}\bigg)(u_1)_{\theta T}-2\sigma\bigg(\frac{w_Y\theta}{w}\bigg)(u_1)_{\theta Y}\notag\\[2mm]
&-\bigg[\bigg(\frac{w_T\theta}{w}\bigg)^2-\sigma\bigg(\frac{w_Y\theta}{w}\bigg)^2\bigg](u_0)_{\theta \theta}\notag\\[2mm]
&-\frac{w_{TT}-\sigma w_{YY}}{w}\theta(u_0)_\theta+\frac{1}{2}\sin(u_0)u_1^2\,,\label{second-order}
\end{align}
where we have introduced the linear operator
\[
{\bf L} = (\xi_T^2-\sigma\xi_Y^2-1)\partial_{\theta \theta}+\cos(u_0)\,.
\]
By differentiation of the first equation in this system with respect to $\theta$,
 it is easy to verify that $(u_0)_\theta \in$ ker$({\bf L})$.
Assuming that this kernel of this operator is one-dimensional, thus being
 spanned by $(u_0)_\theta$, we may deduce the the right hand side of inhomogeneous equations of the form
\[
{\bf L}F = G\,,
\]
lie in the range of the operator ${\bf L}$ whenever one has the inner product condition
\begin{equation}\label{solvability-2}
\int_{-\infty}^\infty (u_0)_\theta G \ d\theta = 0\,.
\end{equation}

Solving this system of equations admits the equations governing the width $w$
 and kink position $\xi$. The first equation of the system, (\ref{zero-order}), 
 defines the solution $u_0$ and admits the (leading order)
 relation between $w$ and $\xi$,
\[
w^2(1-\xi_T^2-\sigma\xi_Y^2) = 1\,.
\]
This is the eikonal equation for the problem, and recovers the leading order 
terms, in the sense of the slow scales introduced for this multiple scales analysis, 
of the $w$-variation of the Lagrangian~(\ref{eq7}).
The equation at the next order, appealing to the solvability condition~(\ref{solvability-2}), 
may be resolved so long as
\begin{equation}\label{cons-law}
(w\xi_T)-\sigma (w\xi_Y)_Y = 0\,,
\end{equation}
thus recovering Eq.~(\ref{eq9}).  This is enough to close the system, meaning that further orders need not be 
considered for a leading order version of the evolution equations for the front parameters.
This approach is therefore seen to recover the leading order
(dispersionless) dynamics of the Lagrangian approach of Sec.~\ref{sec:KAK}), as is expected from
the geometric optics scalings. 

To recover the full set of dynamics predicted by the Lagrangian approach~(\ref{eq7}), one must
reconstitute the analysis (that is, treat all orders of the analysis together) and consider 
its inner product with the generalized kernel element~\cite{kivshar-pelin,kaup}.  
This is equivalent to imposing that the conservation law~(\ref{cons-law}) remains 
unchanged by the presence of $u_1$ to the order of the analysis. 
This element is given by 
\[
\frac{\theta}{w} (u_0)_\theta \equiv \frac{\partial u_0}{\partial w}\,,
\]
and satisfies the generalized eigenvalue problem
\[
{\bf L}\frac{\partial u_0}{\partial w}= 2w(1-\xi_T^2+\sigma \xi_Y^2)(u_0)_{\theta \theta}\,,
\]
a system which can be generated by differentiating Eq.~(\ref{zero-order}) with respect to $w$.
The inner product of this eigenfunction with the reconstituted system of equations gives
\[
\begin{split}
\frac{1}{w^2}\bigg[ &w^2(1-\xi_T^2-\sigma\xi_Y^2)-1\\&+\frac{\varepsilon^2\pi^2}{6}\bigg(\frac{w_{TT}-\sigma w_{YY}}{w}-\frac{w_T^2-\sigma w_Y^2}{4w^2} \bigg) \bigg] = 0\,,
\end{split}
\]
which is equivalent to Eq.~(\ref{eq8}) with the geometric optics scalings imposed. 
It also highlights that the Lagrangian approach of the main paper provides a more accurate 
theory for the evolution of the front parameters, as is the case in the closely related 
Whitham modulation theory~\cite{Infeld-Rowlands,wlnlw}.

\section{Kink stripe stability and dynamics in the hyperbolic dispersion case}
\label{appendix_hyper}

\begin{figure}[t] 
\begin{center}
\includegraphics[width=0.90\columnwidth]{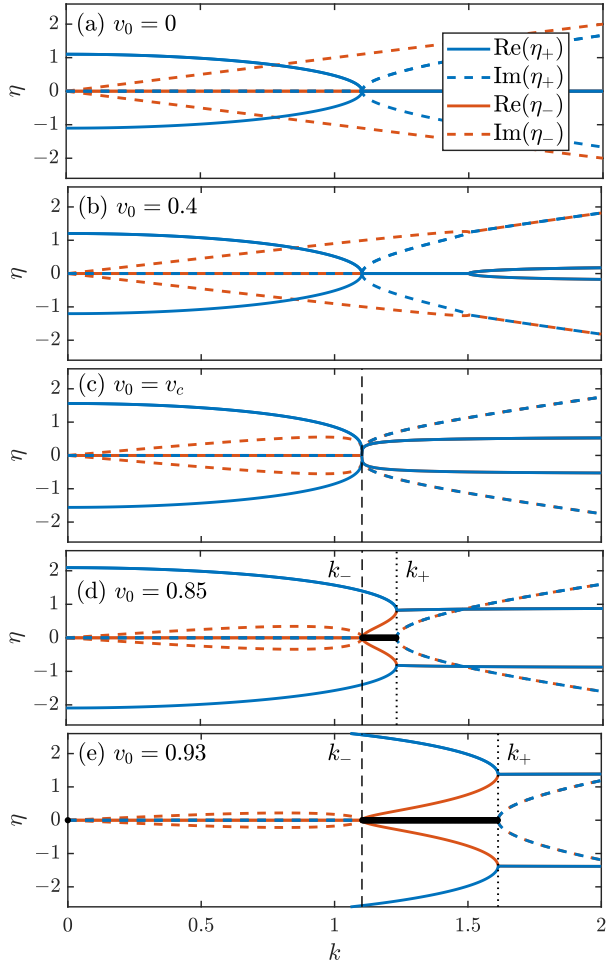}
\caption{(Color online)
Wavenumber stability spectrum for a straight kink moving at
velocity $v_0$ in the hyperbolic $\sigma=-1$ case for the different
velocities indicated in the panels.
For $ v_c=1/\sqrt{2}\leq v_0\leq 1$ wavenumbers in the band $k_-\leq k\leq k_+$
(see black thick line) are (neutrally) stable.
}
\label{fig:unstable_hyper}
\end{center}
\end{figure}

\begin{figure}[t] 
\begin{center}
\includegraphics[width=0.85\columnwidth]{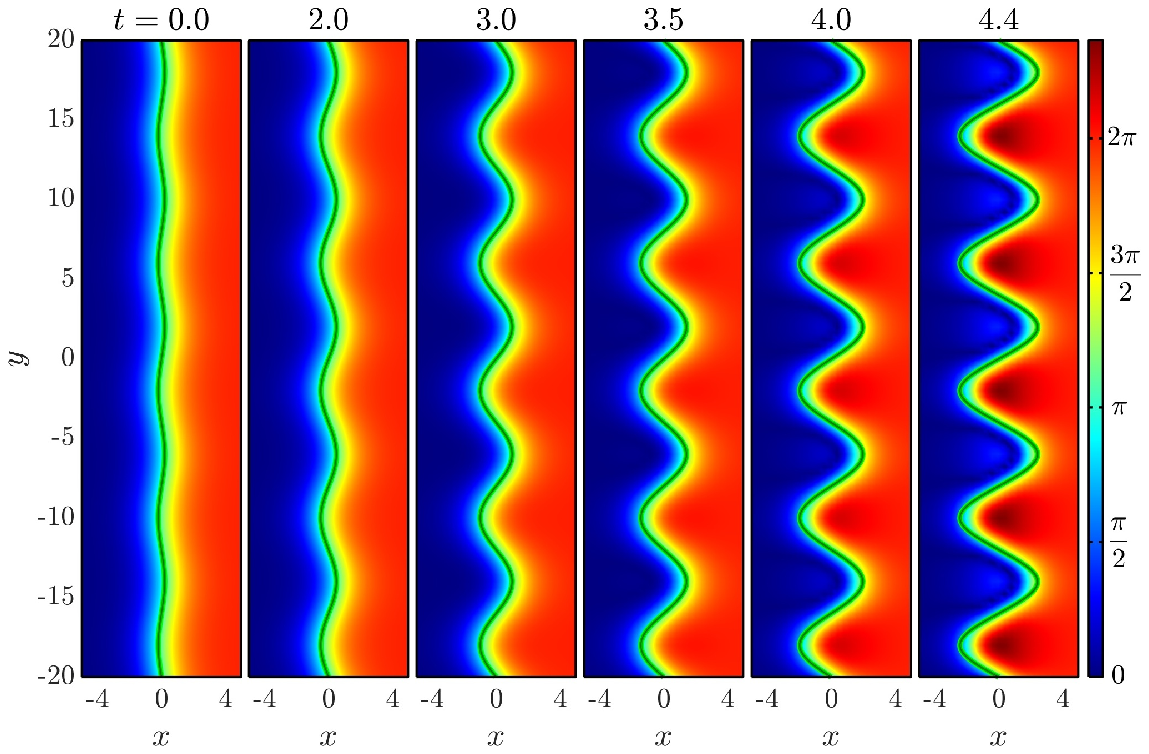}
\\[1.0ex]
\includegraphics[width=0.85\columnwidth]{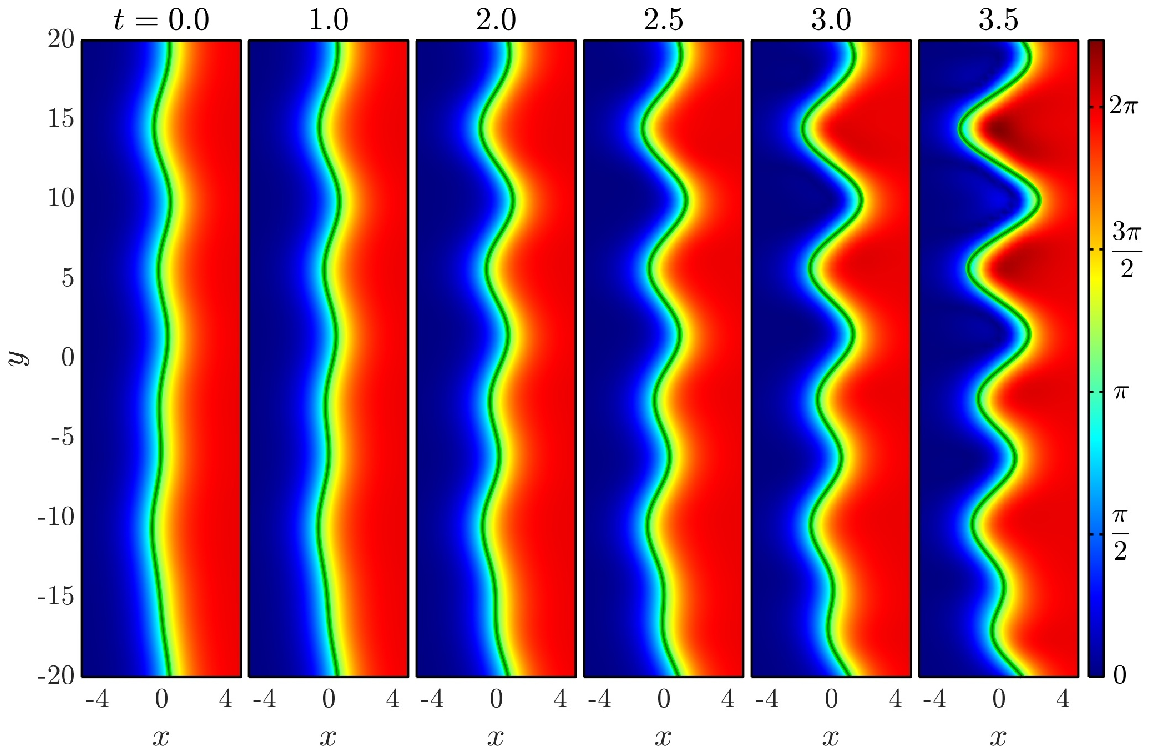}
\caption{(Color online)
Unstable dynamics for a single sG kink stripe in the hyperbolic ($\sigma=-1$) case.
The top row of panels depicts the dynamics for a (slightly) perturbed kink such
that its initial position given by $\xi_0(y) = \varepsilon\sin(2\pi n/\ell)$
where the transverse perturbation mode $n=5$ with amplitude $\varepsilon=0.2$ has 
been excited on the domain $(x,y)\in[-\ell,\ell]\times[-\ell,\ell]$ with $\ell=20$.
The bottom row of panels depicts a similar case where the first 5 transverse
perturbation modes have been perturbed.
The full sG dynamics of Eq.~(\ref{eq:2DsG}) with $\sigma=-1$ 
is depicted by the background color at the indicated times.
The corresponding results of the reduced VA model of
Eqs.~(\ref{eq8}) and (\ref{eq9})
are overlaid by the (green) curves.
Due to the instability in the hyperbolic ($\sigma=-1$) case, shortly after
the last times depicted, the kink stripe breaks down and creates positive
and negatives humps that eventually blow up.
A movie depicting the full stripe dynamics and the reduced VA model
is included in the supplemental material; see movies 
{\tt 1stripe-hyperbolic-mode1.gif} and  {\tt 1stripe-hyperbolic-mode15.gif}.
}
\label{fig:VAhyper}
\end{center}
\end{figure}

Here we study the stability and dynamics for kink stripes in the hyperbolic dispersion
($\sigma=-1$) regime. The hyperbolic nature of the dispersion renders the stripe
generically unstable.
In fact, defining $k_-\equiv2\sqrt{3}/\pi$ and  
$k_+\equiv\sqrt{3}/(\pi v_0\sqrt{1-v_0^2})$, it is straightforward to prove that
all wavenumbers with $k<k_-$ or $k>k_+$ are unstable. Note that if $v_0\rightarrow 1$ then 
the right wavenumber threshold $k_+ \rightarrow \infty$, and 
$k_-=k_+$ yields the critical velocity $v_c=1/\sqrt{2}$. Thus, for $v_c \leq v_0 \leq 1$
there exists a stability band of wavenumbers on the interval $k_-\leq k \leq k_+$.
Figure~\ref{fig:unstable_hyper} shows the stability frequency spectrum for
different values of the stripe's speed $v_0$. Note the stability band (depicted
by the thick black line) between $k_-$ (vertical dashed line) and $k_+$
(vertical dotted line).
Also note that the instability growth rate (Im$(\eta_-)$) for $v_0\rightarrow 1$ corresponding
to $k<k_-$ (see red dashed curves in panel (e) of Fig.~\ref{fig:unstable_hyper}) 
tends to zero and that $k_+ \rightarrow \infty$. Therefore, for velocities 
close to one, the stripe will be very weakly unstable for short
wavenumbers/long wavelengths.

Figure~\ref{fig:VAhyper} depicts a comparison between the full sG dynamics of 
Eqs~(\ref{eq:2DsG}) and the corresponding VA model of Eqs.~(\ref{eq8}) and 
(\ref{eq9}) for $\sigma=-1$.
The figure suggests that the reduced VA dynamics follows
very closely the evolution of the kink stripe in the full model. 
It should be noted that in this hyperbolic case ($\sigma=-1$), the
stripes' undulations rapidly (exponentially) grow and thus the stripes
eventually break up, at which point the VA reduction is not applicable.
Integrating further the full dynamics, one observes that the stripes 
break leaving behind a series of positive and negative humps that, 
in turn, keep growing.

\section{An alternative approach to the transverse stability of a kink stripe}
\label{appendix2}

Here we are interested in providing an alternative
approach to the transverse dynamics of the kink solution, 
as given by Eq.~(\ref{eq:sGkink1D}), with $s=+1$; the analysis for the antikink 
solution, with $s=-1$, is obviously similar to the one with $s=+1$.
%
%

Our approach starts  with the use of the following perturbation ansatz:
\begin{equation}
\label{pert_an}
u(x,t,T_{i},Y_{i})=u_{0}(\xi)+\sum_{i=1}^{n}\epsilon^{i}u_{i}(\xi,T_{i},Y_{i}), 
\end{equation}
where $u_0(\xi)$ is the kink soliton solution, 
\begin{eqnarray}
u_{0}(\xi)&=&4\tan^{-1} \exp\left(\xi \right),
\label{eq:sGkink1D_t} 
\\ 
\notag
\xi &=& w(x-vt-x_0(T_{i},Y_{i})),
\end{eqnarray}
with $x_0(T_{i},Y_{i})$ being the kink center, which is 
assumed to depend on the slow scales $T_{i}=\epsilon^{i}t$ and $Y_{i}=\epsilon^{i}y$,   
where $0<\epsilon\ll1$ is a formal small parameter. On the other hand, the unknown fields 
$u_{i}(\xi,T_{i},Y_{i})$ represent the corrections to the kink shape due to the presence of 
the transverse perturbation, and satisfy the homogeneous boundary condition 
$u_{i}(\xi,T_{i},Y_{i}) \rightarrow 0$, as $\xi \rightarrow \pm \infty$. 

Substituting the perturbation expansion~(\ref{pert_an}) into Eq.~(\ref{eq:2DsG}) 
we obtain the following results. 
First, at $\mathcal{O}(\epsilon^{0})$, we obtain the homogeneous equation:
\begin{equation}
\label{eq:pert0}
(v^{2}-1)\partial_{\xi}^2u_{0}+\sin(u_{0})=0,
\end{equation}
which leads to the kink solution given by Eq.~(\ref{eq:sGkink1D_t}).
At the next orders of approximation, the equations become inhomogeneous, with the 
inhomogeneous parts depending on the derivatives of $x_0$ with respect 
to the slow variables. To be more specific, the resulting equations at orders 
$\mathcal{O}(\epsilon^{i})$ for $i=1,2,3,\ldots$, take the following form:
\begin{equation}
\label{eq:perti}
Lu_{i}=F_{i},
\end{equation}
where the linear operator $L$ (which is the same for every order) is given by 
$L=(v^{2}-1)\partial_{\xi}^2+\cos(u_{0})$, namely the linearization
operator around the kink stripe in the co-traveling frame, while the inhomogeneous parts $F_{i}$, 
up to third order, are given by the following expressions:

\begin{eqnarray}
F_{1}=&-&2v x_{0T_1} u_{0\xi\xi},
\nonumber
\\
F_{2}=&-&2v x_{0T_2} u_{0\xi\xi}- \sigma x_{0Y_1Y_1}u_{0\xi}+x_{0T_1T_1}u_{0\xi}
\nonumber\\
&+& \sigma x_{0Y_1}^{2} u_{0\xi\xi}- x_{0T_1}^{2} u_{0\xi\xi}+ 2vu_{1\xi T_1}
\nonumber\\
&-&2v x_{0T_1} u_{1\xi\xi}+\frac{1}{2}u_{1}^{2}\sin(u_{0}),
\nonumber
\\
F_{3}=&-&2x_{0T_1}x_{0T_2}u_{0\xi\xi}+ 2 \sigma x_{0Y_1}x_{0Y_2}u_{0\xi\xi}
\nonumber\\
&-&2 \sigma x_{0Y_1Y_2}u_{0\xi} + 2 x_{0T_1T_2}u_{0\xi} + \sigma u_{1Y_1Y_1}
\nonumber\\
&-&u_{1T_1T_1}- \sigma x_{0Y_1Y_1}u_{1\xi}+x_{0T_1T_1}u_{1\xi}
\nonumber\\
&-&2 \sigma x_{0Y_1}u_{1\xi Y_1} + 2x_{0T_1}u_{1\xi T_1} + 2vu_{1\xi T_2}
\nonumber\\
&+&\sigma  x_{0Y_1}^{2} u_{1\xi\xi}- x_{0T_1}^{2}u_{1\xi\xi} - 2vx_{0T_2}u_{1\xi\xi}
\nonumber\\
&+&2vu_{2\xi T_1} - 2vx_{0T_1}u_{2\xi\xi} + \sin(u_{0})u_{1}u_{2}
\nonumber\\
&+& \frac{1}{6} \cos(u_{0})u_{1}^{3}.
\nonumber
\end{eqnarray}
To proceed further, it is useful to make a few observations. First, differentiating Eq.~(\ref{eq:pert0}) with respect to $\xi$, one obtains the homogeneous solution $u_h$ of Eqs.~(\ref{eq:perti}), namely $Lu_{0\xi}=0$, implying that $u_h$ is of the form:
\begin{equation}
\nonumber
u_{h}=u_{0\xi}.
\end{equation}
This is naturally a byproduct of the spatial homogeneity of the model,
leading to its translational invariance.
Second, having found the above homogeneous solution, we may derive the solvability conditions 
of the full inhomogeneous problem, Eq.~(\ref{eq:perti}).
To do this, we multiply both sides of Eqs.~(\ref{eq:perti}) by the homogeneous solution
$u_{0\xi}$ and integrate with respect to $\xi$ from $-\infty$ to $+\infty$. Then, 
using
the Hermitian nature of the operator $L$,
as well as the definition of the homogeneous solution ($Lu_{h}=0$), we obtain the solvability conditions in the form of 
the following integral relations:
\begin{equation}
\int_{-\infty}^{\infty}F_{i}u_{h} d\xi=0.
\label{solvability}
\end{equation}
Here, it is important to notice that the above solvability conditions will lead to evolution 
equations for the soliton center $x_{0}$, which will provide the necessary information for 
characterizing the stability of kinks in the 2D space. In addition, the general solutions of  
the ordinary differential equations (ODEs)~(\ref{eq:perti}), which can be used to construct an approximate solution to Eq.~(\ref{eq:2DsG}), can be found by means of the reduction of order 
method; these solutions take the form:
\begin{eqnarray}
u_{i}&=&u_{0\xi} \Bigg[ \int \frac{1}{u_{0\xi}^{2}}
\left(\int\frac{1}{(v^2-1)}u_{0\xi} F_{i}d\xi\right)d\xi
\nonumber \\
&&+\int\frac{A_i(T_i,Y_i)}{u_{0\xi}^2}d\xi + B_i(T_i,Y_i)\Bigg],
\label{ansol}
\end{eqnarray}
where $A_i(T_i,Y_i)$ and $B_i(T_i,Y_i)$ appear due to integration with respect to $\xi$, 
and depend on the slow variables $T_i=\epsilon^{i}t,Y_i=\epsilon^{i} y$. 
Note that, since $B_i(T_i,Y_i)$ can be absorbed in the kink center $x_0$, can be set to zero 
without loss of generality. In addition, it can be seen that $A_i(T_i,Y_i) = 0$ so that 
$u_i$ satisfy the homogeneous boundary conditions. 
Below, for simplicity, we will present results for both the solutions $u_{i}$ and 
the evolution equations for the kink center $x_{0}$, valid up to $\mathcal{O}(\epsilon^{3})$. 
 
At $\mathcal{O}(\epsilon^{1})$ the solvability is satisfied identically, because the 
integrand in Eq.~(\ref{solvability}) is an odd function of $\xi$. Furthermore, 
using Eq.~(\ref{ansol}), we find that the solution $u_{1}$ reads: 
\begin{equation}
u_{1}=\frac{2v}{(1-v^2)^{3/2}}x_{0T_1}\xi\sech\left(\frac{\xi}{\sqrt{1-v^2}}\right).
\label{u_1}
\end{equation}

At $\mathcal{O}(\epsilon^{2})$ the solvability condition in Eq.~(\ref{solvability}) leads 
to the following evolution equation for $x_0$:
\begin{equation}
x_{0T_1 T_1} - \sigma (1-v^2)x_{0Y_1Y_1}=0.
\label{eq:pert1}
\end{equation}
Here, it is important to note that Eq.~(\ref{eq:pert1}) is either elliptic, 
of the Laplace type (for $\sigma=-1$), or hyperbolic, of the form of the usual 2nd-order 
wave equation (for $\sigma=+1$). This means that, in the former case with $\sigma=-1$, 
the solution will grow exponentially, i.e., $x_0 \propto \exp(cK T_1)$, where $c^2=1-v^2$ 
and $K$ is the wavenumber of the transverse perturbation.  
In this case [corresponding to the hyperbolic form of the spatial
operator in Eq.~(\ref{eq:2DsG})], the kink is 
transversely unstable. On the other hand, if $\sigma=+1$ [corresponding to the elliptic form of 
the spatial operator in Eq.~(\ref{eq:2DsG})], $x_0$ never grows and the kink solution is transversely stable, 
up to this order of approximation, i.e., at the scales $T_1=\epsilon t$ and  $Y_1=\epsilon y$. 
In such a case, the general solution of Eq.~(\ref{eq:pert1}) is of the
classical D'Alembert form:
\begin{equation}
x_{0}=\Phi(Y_1-cT_1,~T_2,Y_2)+\Psi(Y_1+cT_1,~T_2,Y_2),
\label{solpert1}
\end{equation}
where the functions $\Phi$ and $\Psi$ can be found from the initial data.  
Next, at the same order of approximation [$\mathcal{O}(\epsilon^2)$], we proceed to 
calculate the second order term, $u_2$, in the perturbation expansion. Using again 
Eq.~(\ref{ansol}) we obtain:
\begin{eqnarray}
u_{2}&=&\frac{1}{(1-v^2)^{5/2}}\xi \sech\left(\frac{\xi}{\sqrt{1-v^2}}\right) 
\label{u_2}
\\[1.0ex]
&&\times  \Bigg[(1-v^2)\left(
2v x_{0T_2} - \sigma v^2 \xi x_{0Y_1Y_1} -\sigma x_{0Y_1}^{2} \right)
\nonumber\\
\nonumber
&&+ (1+2v^2)x_{0T_1}^{2}+\frac{v^2x_{0T_1}^{2}}{\sqrt{1-v^2}}
\xi \tanh\left(\frac{\xi}{\sqrt{1-v^2}}\right)\Bigg].
\end{eqnarray}
Focusing on the stable case of $\sigma=+1$, we may keep only linear terms in $x_{0}$, 
corresponding to the case of a sufficiently small perturbation such that 
$|x_0| \ll 1$, and simplify Eq.~(\ref{u_2}) to obtain the approximate expression: 
\begin{eqnarray}
u_{2}&\approx&\frac{2v}{(1-v^2)^{3/2}} \left(x_{0T_2}- \frac{1}{2}v\xi x_{0Y_1Y_1}\right)
\nonumber\\
&&\times~
\xi \sech\left(\frac{\xi}{\sqrt{1-v^2}}\right).
\label{u_21}
\end{eqnarray}

\begin{figure}[t] 
\begin{center}
\includegraphics[width=0.90\columnwidth]{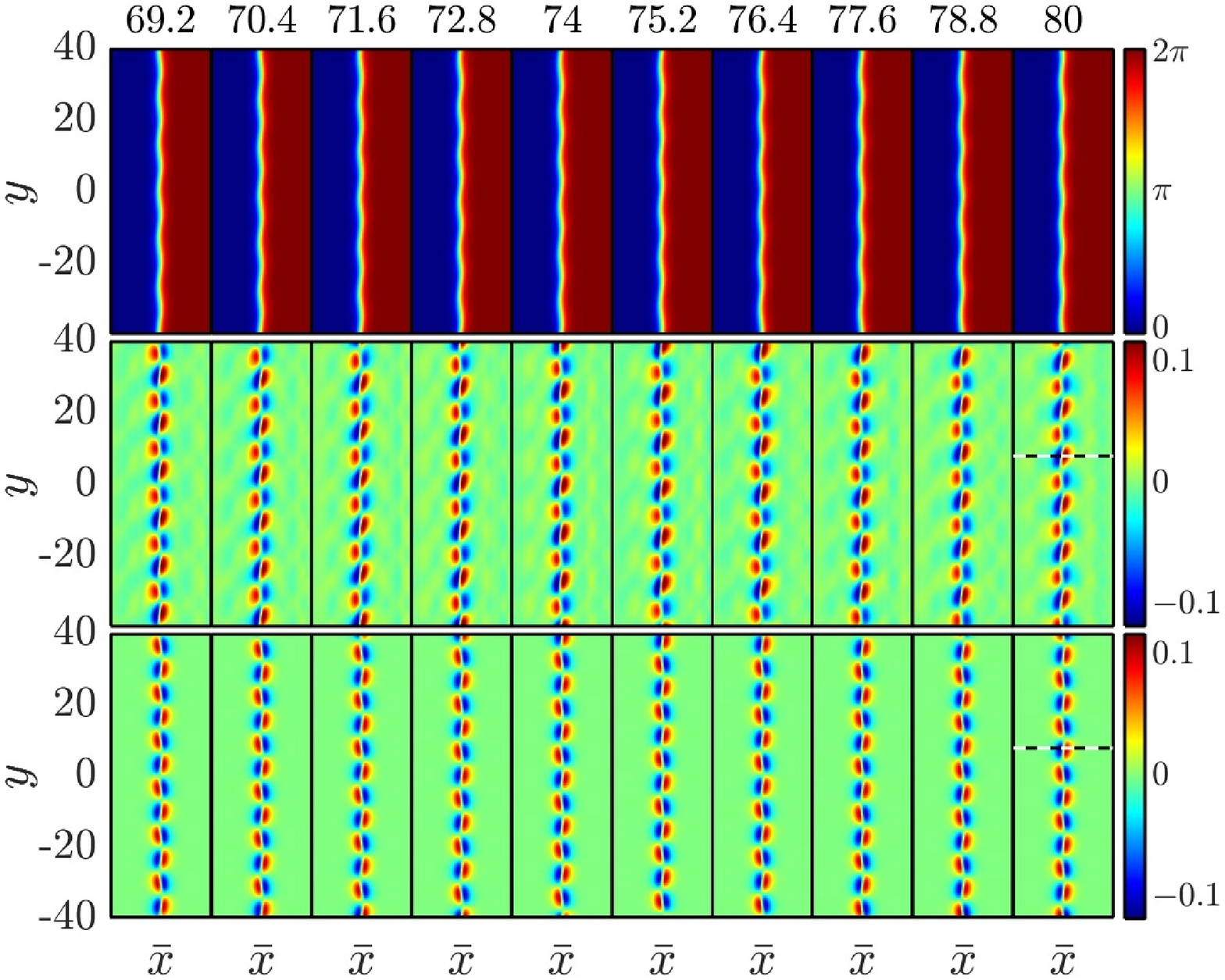}
\\[2.0ex]
\includegraphics[width=0.80\columnwidth]{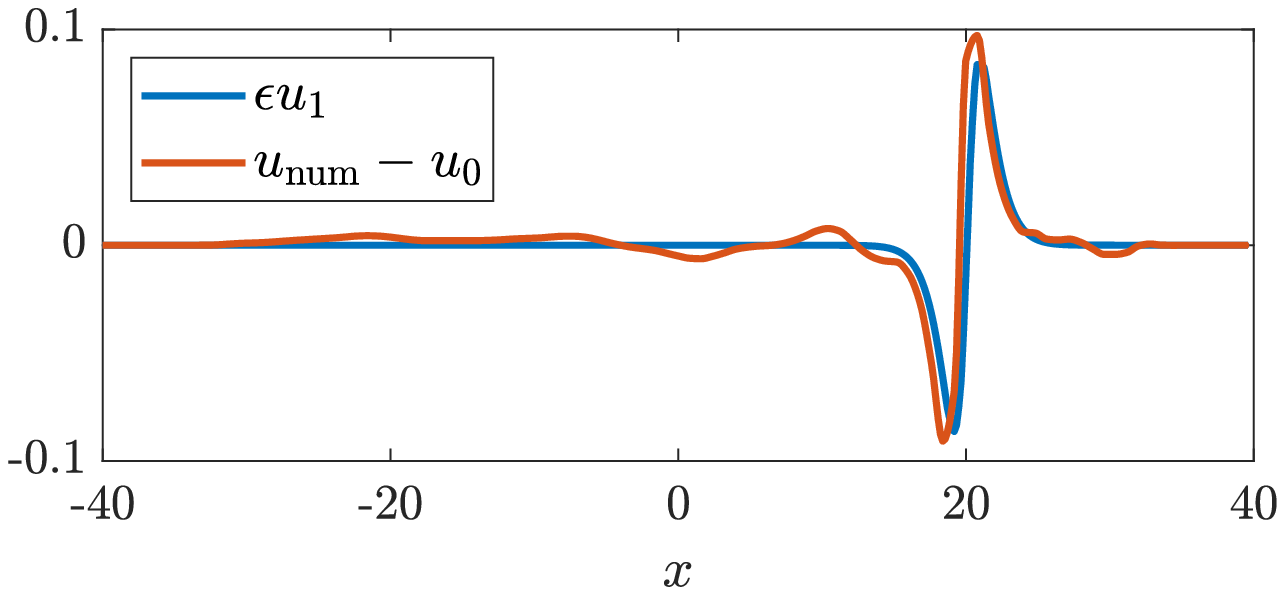}
\caption{(Color online)
Validating the perturbation theory for transverse perturbations of a
traveling kink stripe.
Top row of panels: Evolution of a perturbed kink stripe at the indicated times. 
The initial condition for the stripe 
corresponds to $u(x,y,t=0)=u_0+\epsilon u_1$ where $u_0$ is given by 
Eq.~(\ref{eq:sGkink1D_t}) with $x_0=d\cos(n\pi y/\ell)$ with 
$v=0.5$, $d=0.25$, $n=6$ and the perturbation $u_1$ is defined in Eq.~(\ref{u_1})  
for $\epsilon=0.1$ on the domain $[-\ell,\ell]\times[-\ell,\ell]$ with $\ell=40$.
Second row of panels: Evolution of the perturbation $u_1$ extracted by subtracting the 
unperturbed stripe~(\ref{eq:sGkink1D_t}) from the full numerics of $u(x,y,t)$
shown in the top row of panels.
Third row of panels: Corresponding evolution of the perturbation according to the
perturbation theory to order $\mathcal{O}(\epsilon)$.
The $x$-axis is taken in the co-traveling frame $\bar{x}=x-vt$ that
follows the kink stripe movement to the right at velocity $v$.
Bottom panel: Transverse cut for $y=8$ at $t=80$ (see vertical lines in the
last panel of the second and third rows) comparing the perturbation extracted from
full numerics ($u_{\rm num}-u_0$) and the perturbation from the
perturbation theory to order $\mathcal{O}(\epsilon)$.
%
}
\label{fig:perturbation_theo_stripe}
\end{center}
\end{figure}

Figure~\ref{fig:perturbation_theo_stripe} depicts an example validating the
perturbation approach to order $\mathcal{O}(\epsilon)$. For this example,
a traveling undulation of amplitude 0.25 is mounted  on top of a
kink stripe traveling at velocity $v=0.5$. The full numerical solution
on the co-moving reference frame $\bar{x}=x-vt$ is depicted in the first 
row of panels in Fig.~\ref{fig:perturbation_theo_stripe}. The second and third
row of panels in the figure depict, respectively, the numerically extracted
perturbation ---by subtracting the unperturbed shifted stripe~(\ref{eq:sGkink1D_t}) 
from the full numerical solution--- and the perturbation predicted by
the perturbation theory to order $\mathcal{O}(\epsilon)$.
As these panels suggest, the perturbation theory accurately captures
the evolution of the perturbation on top of the traveling and
undulating kink stripe.
We note that adding the perturbation terms to order $\mathcal{O}(\epsilon^2)$
did not visibly change the results presented in the figure.
Nevertheless, it is important to highlight that the following quantitative
differences arise between the second and third row of panels of
Fig.~\ref{fig:perturbation_theo_stripe}.
The former features small amplitude dispersive wavepackets (radiation) which the
theoretical approach obviously cannot capture (by construction). These appear
as a faint (yellow-colored) halo in the second row panels of the figure.
The better visualize these radiation effects, we depict in
the bottom panel of the figure a cut for $y=8$ at $t=80$ 
(see vertical lines in the last panel of the second and third rows)
for the numerically extracted perturbation and the one predicted
by the perturbation theory to order $\mathcal{O}(\epsilon)$.
This last panel evidences the fact that, as an astute reader would
have noted, while the wavenumber
and overall spatial behavior of the perturbation is accurately captured,
details of the match reveal quantitative discrepancies as regards the amplitude
of the relevant perturbation of the order of $\approx 20\%$.

It is also possible to further continue the perturbation expansion
procedure at the next order of approximation, namely 
at $\mathcal{O}(\epsilon^{3})$. Indeed, at this order, the solvability 
condition~(\ref{solvability}) leads to the following equation for $x_{0}$:
\begin{equation}
x_{0T_1T_2}-\sigma (1-v^2)x_{0Y_1Y_2} = 0,
\label{eq:pert2}
\end{equation}
which has a form similar to that of Eq.~(\ref{eq:pert1}), but at scales $T_2 = \epsilon^2 t$ 
and $Y_2 = \epsilon^2 y$. Hence, one can infer that the kink is stable 
(at these scales) only for $\sigma=+1$. In this case, we may find a solution of Eq.~(\ref{eq:pert2}) 
employing the results obtained at $\mathcal{O}(\epsilon^2)$. Indeed, choosing 
(without loss of generality) the right-going wave $\Phi(Y_1-cT_1,~T_2,Y_2)$ of 
Eq.~(\ref{solpert1}), it is straightforward to derive from Eq.~(\ref{eq:pert2}) the following 
transport equation for the field $\Phi(Y_1-cT_1,~T_2,Y_2)$:
\begin{equation}
\Phi_{T_2}+c\,\Phi_{Y_2}=0,
\label{eq:phi}
\end{equation}
which has the solution $\Phi=\Phi(Y_1-cT_1,~Y_2-cT_2)$. Thus, for $\sigma=+1$, the kink 
is transversely stable not only for the scales $T_1=\epsilon t$ and $Y_1=\epsilon y$, but also for 
$T_2=\epsilon^{2} t$ and  $Y_2=\epsilon^{2} y$. 

At the same order, $\mathcal{O}(\epsilon^3)$, one may again use Eq.~(\ref{ansol}), and keeping 
linear terms in $x_{0}$ as above,  obtain the following expression for $u_{3}$ 
(for $\sigma=+1$):
\begin{eqnarray}
u_{3}&=&\frac{v^2}{3(1-v^2)^{5/2}} \left[v\xi x_{0T_1Y_1Y_1} -6(1-v^2)x_{0Y_1Y_2} \right] 
\nonumber\\
&&\times ~ \xi^{2} \sech\left(\frac{\xi}{\sqrt{1-v^2}}\right).
\label{u_3}
\end{eqnarray}

In principle, the above analysis can be continued to higher orders of approximation, namely at 
$\mathcal{O}(\epsilon^{4})$, $\mathcal{O}(\epsilon^{5})$ and so on. We surmise that our 
main result, namely that the kink (or the antikink) solution of the
elliptic (in its spatial operator)
sG equation is stable, will remain the same. Our analysis suggests that the kink 
center $x_{0}$ will always satisfy a transport equation for the right- or the left-going wave, 
remaining in this way always bounded. In addition, the above analysis offers an approximate analytical solution [valid up to $\mathcal{O}(\epsilon^{3})$] of the full 2D sG model, of the form 
$u\approx \sum_{n =0}^{3} \epsilon^n u_n,$ 
where $u_0$ is the unperturbed soliton given by Eq.~(\ref{eq:sGkink1D_t}), while $u_1$, 
$u_2$ and $u_3$ are respectively given by Eqs.~(\ref{u_1}), (\ref{u_21}) and (\ref{u_3}).  

\section*{Acknowledgements}
%
This material is based upon work supported by the US National Science Foundation
under Grants PHY-1602994, DMS-1809074, PHY-2110030 (P.G.K.).
R.C.G.~gratefully acknowledges support from the US National Science Foundation
under Grants PHY-1603058 and PHY-2110038.
%


\section*{References}

\end{document}